\documentclass[apj]{emulateapj}
\usepackage{apjfonts}

\gdef\h50min{$h_{50}^{-1}$}
\gdef\kms{km\,s$^{-1}$}

\gdef\3727{[O\,{\sc ii}]\,3727\,\AA}
\gdef\5007{$\lambda \lambda 5007$\,[O\,{\sc iii}]}

\gdef\4ang{4000\,\AA}
\gdef\clusa{3C\,295}
\gdef\clusb{CL\,0016+16}
\gdef\clusc{CL\,1601+42}
\gdef\clusbs{CL\,0016}
\gdef\cluscs{CL\,1601}
\gdef\msun{$M_{\odot}$}
\lefthead{van Dokkum \& van der Marel}
\righthead{Star Formation Epoch of Early-Type Galaxies}
\slugcomment{Accepted for publication in the Astrophysical Journal}
\begin{document}

\title{The Star Formation Epoch of the most Massive Early-Type
Galaxies\altaffilmark{1,2}}

\author{Pieter~G.~van Dokkum\altaffilmark{3,4}
and
Roeland~P.~van der Marel\altaffilmark{5}
}

\altaffiltext{1}
{Based on observations obtained at the W.\ M.\ Keck Observatory,
which is operated jointly by the California Institute of
Technology and the University of California.}
\altaffiltext{2}
{Based on observations with the NASA/ESA {\em Hubble Space
Telescope}, obtained at the Space Telescope Science Institute, which
is operated by AURA, Inc., under NASA contract NAS 5--26555.}
\altaffiltext{3}{Department of Astronomy, Yale University,
New Haven, CT 06520-8101}
\altaffiltext{4}{Previous address:
California Institute of Technology, MS 105-24, Pasadena, CA 91125}
\altaffiltext{5}{Space Telescope Science Institute, 3700 San Martin
Drive, Baltimore, MD 21218}

\begin{abstract}

We present new spatially-resolved Keck spectroscopy of early-type galaxies
in three galaxy clusters at $z\approx 0.5$. 
In two companion papers (van der Marel \& van Dokkum 2006a,b) we 
construct dynamical models of the galaxies
and compare their modeled $M/L$ ratios and dynamical structure
to local samples.
Here we focus on the fundamental plane (FP) relation, and
combine the kinematics
with structural parameters determined from {\em Hubble
Space Telescope} ({\it HST}) images.
The galaxies obey clear FP
relations, which are offset from the FP of the nearby Coma
cluster due to passive evolution of the stellar populations.  The
$z\approx 0.5$ data are combined with published data for eleven
additional clusters at $0.18\leq z \leq 1.28$, to determine the
evolution of the mean $M/L_B$ ratio of cluster galaxies with masses
$M\gtrsim 10^{11}\,$M$_{\odot}$, as implied by the FP.  We find
$d \log (M/L_B)/dz = -0.555 \pm 0.042$, stronger evolution than
was previously inferred from smaller samples. The observed
evolution depends on the luminosity-weighted mean
age of the stars in the galaxies, the initial mass function (IMF),
selection effects due to progenitor bias, and other
parameters. Assuming a normal IMF but allowing for various other
sources of uncertainty, we find $z_*=2.01^{+0.22}_{-0.17}$ for the
luminosity-weighted mean star formation epoch. The
main uncertainty is the slope of the IMF in the range $1-2\,M_{\odot}$:
we find $z_* = 4.0$ for a top-heavy
IMF with slope $x=0$.  The $M/L_B$ ratios of the cluster galaxies are
compared to those of field early-type
galaxies at $0.32\leq z\leq 1.14$.  Assuming that progenitor bias and
the IMF do not depend on environment we find that the present-day age
of stars in massive field galaxies is
4.1\,\%\,$\pm$\,2.0\,\% ($\approx 0.4$\,Gyr) less than that of
stars in massive
cluster galaxies.
This relatively small age difference is
surprising in the context of expectations from ``standard'' hierarchical galaxy
formation models, and provides a constraint on the physical processes
that are responsible for halting star formation in the progenitors of
today's most massive galaxies. 

\end{abstract}


\section{Introduction}

Recent studies of near-infrared selected samples have
uncovered a large population of red, massive galaxies at $z>2$
(e.g., {Labb{\' e}} {et~al.} 2003; {Franx} {et~al.} 2003; {van Dokkum} {et~al.} 2003; {Daddi} {et~al.} 2004; {Yan} {et~al.} 2004; {Papovich} {et~al.} 2006). Red galaxies
dominate the high-mass end of the mass function at $2<z<3$
({van Dokkum} {et~al.} 2006) and are highly clustered (Daddi et al.\ 2003;
Grazian et al.\ 2006; Quadri et al.\ 2006).
Most of these objects are too faint in the rest-frame ultra-violet
to be selected as Lyman break galaxies
(e.g., {F{\"o}rster Schreiber} {et~al.} 2004; {Reddy} {et~al.} 2005).
The most straightforward interpretation of these newly-found objects is
that they are progenitors of today's massive early-type galaxies.

Interestingly, a large fraction of these galaxies appear
to have very high star formation rates, based on modeling of their
spectral energy distributions, (stacked) X-ray emission,
infrared emission, submm emission, and spectra
(e.g., {F{\"o}rster Schreiber} {et~al.} 2004; {Rubin} {et~al.} 2004; {Daddi} {et~al.} 2004; {Labb{\'e}} {et~al.} 2005; {Knudsen} {et~al.} 2005; {Papovich} {et~al.} 2006; {Webb} {et~al.} 2006). A similar or smaller fraction is best-fit by passively
evolving models with little ongoing star formation
(e.g., {Labb{\'e}} {et~al.} 2005; {Papovich} {et~al.} 2006; {Kriek} {et~al.} 2006b). These results suggest that by
$z\sim 2.5$ we are entering
the star formation epoch of massive early-type galaxies.

An important question is whether the properties of early-type
galaxies at intermediate redshift are consistent with this interpretation.
Furthermore, detailed studies of early-type galaxies in different
environments may identify descendants of particular populations of
massive galaxies at $z>2$. For example, it may be that the passively
evolving galaxies
at $z\sim 2.5$ are progenitors of cluster early-type galaxies whereas
the star-forming galaxies are progenitors of early-type galaxies
in groups.

One of the most sensitive tools for determining the star formation
epoch of early-type galaxies is the redshift evolution of the
fundamental plane (FP)
relation ({Djorgovski} \& {Davis} 1987), as it reflects evolution in the
$M/L$ ratios of galaxies and has very low scatter.
The $z>0$ FP has been studied extensively in the past decade, both in
clusters (e.g., {van Dokkum} \& {Franx} 1996; {van Dokkum} {et~al.} 1998a; {Wuyts} {et~al.} 2004; {Fritz} {et~al.} 2005; {Holden} {et~al.} 2005;
{J{\o}rgensen} {et~al.} 2006) and
in the field (e.g., {Treu} {et~al.} 1999, 2002, 2005a; {van Dokkum} {et~al.} 2001; Rusin et al.\ 2003; van de Ven et al.\ 2003;
{van der Wel} {et~al.} 2004, 2005; {di Serego Alighieri} {et~al.} 2005).
The results from these papers can be summarized
as follows: 1) the $M/L$ ratios
of the most massive cluster and field early-type galaxies evolve slowly
and regularly, indicating early formation of their stars; and 2)
there is evidence that low mass galaxies evolve faster than high mass
galaxies, both in the field and in clusters, suggesting that they
have younger stellar populations. The interpretation of the first
result is complicated by the small number of clusters that have been
studied so far, uncertainties in the initial mass function (IMF),
and by selection effects due to progenitor bias: if the morphologies
of galaxies change with time the sample of high-redshift early-type
galaxies is only a subset of the sample of nearby early-type galaxies,
leading to biased age estimates (see {van Dokkum} \& {Franx} 2001).
The interpretation of the second result is complicated by the fact that
large corrections for selection effects need to be made
(see {Treu} {et~al.} 2005a, 2005b; {van der Wel} {et~al.} 2005), and that
galaxies are assumed to undergo no structural or dynamical changes with
redshift.

The goal of the present paper is to better constrain the star formation
epoch of massive field and cluster galaxies, using the FP.
We present
spatially resolved spectroscopic data from Keck and high-quality
photometric data from the {\em Hubble Space Telescope} (HST)
for a sample of early-type
galaxies in three clusters at $z\approx 0.5$.
The FPs of the three clusters are presented
and discussed, and 
combined with literature
data on the nearby Coma cluster, a sample of nearby galaxies drawn
from the Sloan Digital Sky Survey (SDSS), and eleven clusters at
$0.18\leq z \leq 1.28$. The evolution of the $M/L$ ratio derived
from this large sample is used to constrain
the star formation epoch of massive cluster galaxies.
We combine the cluster sample with
recently published samples of field galaxies, and determine the
age difference between massive field and cluster galaxies using a
self-consistent modeling approach which is different from previous
studies. In two companion papers
(van der Marel and van Dokkum 2006a,b; hereafter vdMvD06a,b)
we utilize our spatially-resolved
data to construct detailed dynamical models for the sample galaxies.
The results are compared to those for local samples, providing an
independent method of deriving $M/L$ ratio evolution, and
allowing us to validate the many assumptions that enter into
analyses based on the FP. The dynamical models
also yield a normalized measure of the rotation rate
[akin to $(v/\sigma)^*$] of the sample galaxies, providing a
method to identify S0 galaxies that have been visually misclassified
as ellipticals, based on their kinematics.

The structure of the present
paper is as follows. The analysis of the
spectroscopic and
photometric data for the three $z\approx 0.5$ clusters is described
in \S\,2 and \S\,3.  In \S\,4 the
FPs of the three clusters are presented and discussed. In \S\,5
results for the three clusters are combined with literature data
on 11 additional distant clusters and two local samples,
and the $M/L_B$ evolution is determined from this large sample.
Detailed information on the transformations of all the literature
data to our system are given in an Appendix. The measured evolution
of the $M/L_B$ ratio is interpreted in \S\,6. In this Section
the implications for the star formation epoch of massive cluster galaxies
are discussed, and the cluster data are compared to previously
published data for field early-type galaxies. The main results
are summarized and discussed in \S\,7.
We assume $\Omega_m=0.3$, $\Omega_{\Lambda}=0.7$, and $H_0=71$\,\kms\,Mpc$^{-1}$
where needed.

\section{Spectroscopy}

\subsection{Cluster Selection}

Galaxies were selected from the {Smail} {et~al.} (1997) [MORPHS]
catalogs of the $z\approx 0.5$ galaxy clusters \clusa, \clusb, and \clusc.
These three clusters were selected
based on their visibility at the time of our Keck observations,
and because they are among the most S0-deficient clusters
in the MORPHS sample. The MORPHS sample itself was not selected
according to strict criteria.
Properties of the three clusters are listed in Table 1. They
span a range of a factor of $\sim 17$ in
X-ray luminosity. \clusa\ and \clusc\  have very high measured
velocity dispersions, possibly indicating substructure along the line-of-sight.
The dominant galaxy in \clusa\ is a strong radio source.
All MORPHS clusters have been observed with the {\em HST} WFPC2
camera, and
bright galaxies in the WFPC2 fields
were visually classified by the MORPHS team ({Smail} {et~al.} 1997; {Dressler} {et~al.} 1997).

\begin{small}
\begin{center}
{ {\sc TABLE 1} \\
\sc Cluster Properties}$^a$ \\
\vspace{0.1cm}
\begin{tabular}{lccc}
\hline
\hline
 & \clusa\ & \clusb\ & \clusc \\
\hline
$z$ & 0.460 & 0.546 & 0.538\\
$L_X^b$ & 3.20 & 5.88 & 0.35 \\
$\sigma$ (km\,s$^{-1}$) & 1670 & 1703 & 1166\\
F555W $T_{\rm exp}$ (ks) & --- & 12.6 & --- \\
F702W $T_{\rm exp}$ (ks) & 12.6 & --- & 16.8 \\
F814W $T_{\rm exp}$ (ks) & --- & 16.8 & --- \\
E fraction$^c$ & 0.51 & 0.58 & 0.33 \\
S0 fraction$^c$ & 0.21 & 0.15 & 0.12 \\
S fraction$^c$ & 0.28 & 0.27 & 0.55 \\
\hline
\end{tabular}
\end{center}
{\small
$^a$\,Taken from Smail et al.\ (1997)\\
$^b$\,0.3$-$3.5\,keV; $\times 10^{44} h^{-2}$\,ergs\,s$^{-1}$\\
$^c$\,Morphological fractions from Dressler et al.\ (1997)\\
}
\end{small}

\subsection{Galaxy Selection and Observations}

The sample selection was largely constrained by the geometry of the
Keck Low Resolution Imager and Spectrograph (LRIS) masks. Priority was
given to visually-classified E and E/S0 galaxies with $R_{702}<21.5$
(\clusa\ and \clusc) or $I_{814}<20.8$ (\clusb).  For each cluster
two masks were designed. The individual slits were tilted in order to
optimally align them with the major axes of the galaxies, within the
range allowed by geometrical restrictions. Some galaxies are contained
in both masks; for those objects we used position angles on the sky
that are offset by 90 degrees, or the same position angle if the slit
could be aligned with the major axis in both masks. A large, bright
S0/Sb galaxy was included in one of the \clusa\ masks, to test whether
rotation can be reliably measured for an ``obvious'' disk galaxy.
Remaining space in the multi-slit masks was used to observe candidate
lensed galaxies and random fainter objects in the WFPC2 field.
These secondary ``filler'' objects 
are not discussed in this paper.

The three clusters were observed with LRIS on 2001 18--19
June. Conditions were variable, with part of the first night lost to
clouds. The seeing varied from $0\farcs 7 - 0\farcs 9$
during the run.
For \clusb\ and \clusc\ exposures were obtained for both masks;
for \clusa\ only one mask was observed. The red arm of LRIS was used
with the 900 lines\,mm$^{-1}$ grating blazed at 5500\,\AA.
The $1\farcs 1$ wide slits provide a spectral resolution (as measured from
sky emission lines) $\sigma_{\rm instr} \approx 65$\,\kms.
The observed wavelength range is different for each object (as it depends
on its position in the multi-slit mask), but is typically
3500--4800\,\AA\ in the rest-frame.  Total exposure times were 5400\,s
per mask for \clusa\ and \clusb, and 9000\,s per mask for \clusc.  The
exposures were split into individual 1800\,s exposures to facilitate
cosmic-ray removal. The telescope was not moved in the slit-direction
between successive exposures, as this would have moved the objects out
of the (tilted) slitlets.
Calibration exposures (dome flats, arc lamps) were obtained in daytime.
Dome flats were taken at a range of zenith angles.

\subsection{Reduction}

The reduction followed standard procedures for multi-slit
spectroscopic data (see, e.g., {Kelson} {et~al.} 2000b; {van Dokkum} \& {Stanford} 2003).
Each slitlet in the masks was treated as a separate long slit
spectrum.  Bias was subtracted by fitting a low order polynomial to
the overscan regions. The ccd was read out using two amplifiers; the
bias of each amplifier was fitted separately. Two bad columns were
linearly interpolated.  The dome flats were used to flat field the
data.  Before flat fielding, the dome flat exposures were divided by a
polynomial fit in the $\lambda$ and spatial directions, as the
illumination of the ccd is different from that of the night sky
emission.

\begin{figure*}[t]
\epsfxsize=16.5cm
\epsffile[68 184 486 672]{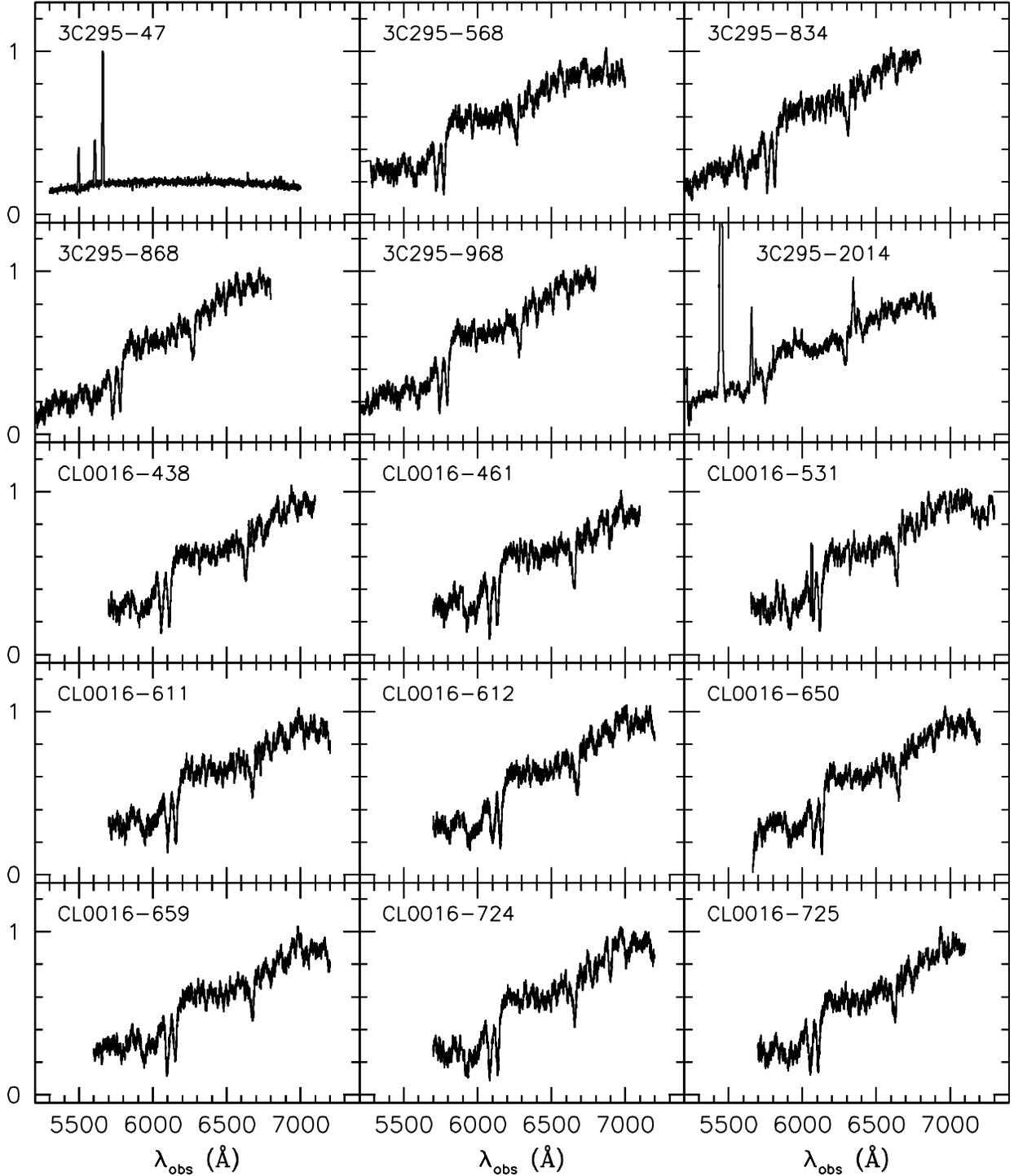}
\caption{\small
Keck spectra of all 27 targeted  galaxies in the three cluster fields.
The spectra are not binned or smoothed, and have
a resolution of $\sigma \approx 65$\,\kms. The vertical scale is
arbitrary, as the spectra have not been calibrated.
The spectroscopy shows
that galaxies 3C\,295--47 ($z=0.1308$)
and CL\,1601--270 ($z=0.5098$) are field galaxies unrelated to the
clusters. These two galaxies are not included in the subsequent
figures and analysis.
\label{specs.plot}}
\end{figure*}

\addtocounter{figure}{-1}

\begin{figure*}[t]
\epsfxsize=16.5cm
\epsffile[68 275 486 672]{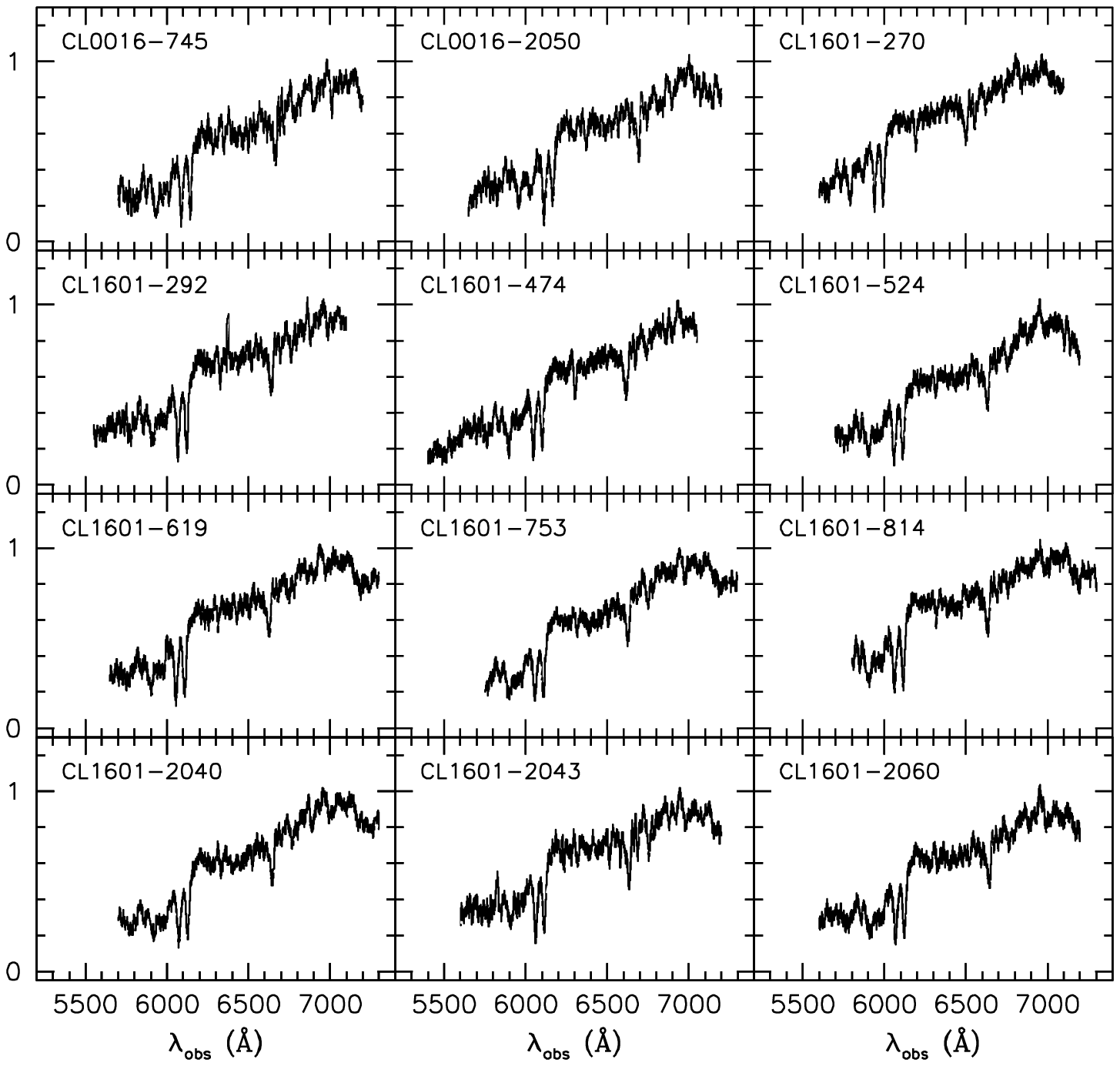}
\caption{\small
(continued)
}
\end{figure*}

Bright cosmic rays were removed from each slitlet in the following
way. A first order approximation of the sky emission lines was
subtracted by fitting a low order polynomial. This step reduced the
intensity of the lines but did not remove them completely because most
slits are tilted. The sky line residuals and the galaxy spectra were
largely removed by subtracting the median of the exposures.  Remaining
residuals were removed by fitting a polynomial, first in the spatial
direction and then in the wavelength direction. This three-step
procedure proved to be much more effective in removing the light of
galaxies and night sky emission than simply subtracting the median.
Cosmic rays were identified by comparing the counts to a noise model
generated from the subtracted median image and the polynomial
fits. Pixels deviating more than 4\,$\sigma$ were flagged as cosmic
ray hits.  

Sky subtraction of tilted slits is difficult, as the standard fitting
procedures require that sky lines are fairly well aligned with
columns. Straightening the sky lines before subtracting them has the
disadvantage of introducing aliasing effects due to the relatively
large pixels of the LRIS ccd, which are very difficult to remove.
More complex procedures have been developed which circumvent these
problems (e.g., {Kelson} 2003); however, as we are working in a
wavelength region mostly blueward of the ubiquitous OH lines and the
galaxies are relatively bright, we found that an iterative approach
provided satisfactory results. At this point in the reduction we
subtracted sky lines by simply fitting a low order polynomial in the
spatial direction, masking the galaxy spectrum and any other objects
in the slit.  While not completely removing the lines, this step
greatly reduces the aliasing effects that occur when resampling the
lines after wavelength calibration.

Wavelength calibration was done using arc lamp exposures, with all
lamps on (Hg, Ne, Ar, Cd, and Zn). Line identification was done
separately for each row, as the solution is a strong function of
position for these tilted slits. The $\lambda \lambda$\,6300.4 O\,{\sc
i} sky line was used to apply small corrections to the zeropoints of
the wavelength solutions. The S-distortion of the spectra was
determined by fitting the wavelength dependence of the position of the
galaxy spectra. Two-dimensional polynomials were fitted to the
measured positions of the sky lines and galaxy spectra, and these were
used to rectify the spectra. In this procedure the spectra are
resampled only once. After rectification, remaining sky line residuals
were removed by fitting a polynomial (masking the galaxy spectra),
and remaining faint cosmic rays were removed with {\sc L.A.Cosmic}
({van Dokkum} 2001).

Spectra of the 27 target objects are shown in Fig.\ \ref{specs.plot}.
They were created by averaging the central five rows ($1\farcs 1$),
and have not been weighted,
smoothed or binned. The sampling is $38$\,\kms\
pixel$^{-1}$, and $\sigma_{\rm instr}\approx 65$\,\kms. The S/N
varies from $\sim 20$ to $\sim 60$ per \AA. Note that the
spectra are not flux-calibrated.

\subsection{Redshifts}

Redshifts were determined using the cross-correlation program {\sc
xcsao} in IRAF or, in one case, from emission lines. A nearby
early-type galaxy was used as the template in the cross-correlation.
The redshifts are listed in Table 2 along with the {Smail} {et~al.} (1997)
morphological classifications.  Of 27 observed galaxies 25 are at the
redshift of the cluster. Of those 25, 24 have an early-type spectrum
without strong emission lines. The exception is \clusa--2014, which
shows broad emission lines superposed on an early-type spectrum. The
presence of AGN features in this object is not surprising as it hosts
the bright radio source 3C\,295, the namesake of the cluster.

The two galaxies with deviant redshifts are \clusa--47 and \cluscs--270.
\clusa--47 is an emission-line object at $z=0.1308$. The lines
are strong and narrow: the \5007\ line has a rest-frame equivalent
width $W_{\lambda}\approx 45$\,\AA, and is
unresolved at our resolution of $\approx 65$\,\kms. The galaxy resembles
the Compact Narrow Emission Line Galaxies described by
{Guzman} {et~al.} (1996). \cluscs--270 is a field early-type
galaxy $\sim 10^4$\,\kms\ removed from the mean recession velocity of
\clusc. These galaxies are not included in the
subsequent Figures and analysis. We note that galaxy
\cluscs--270 is included in the modeling of vdMvD06a,b.

\subsection{Velocity Dispersions}

Internal velocity dispersions were determined from direct fits of the
galaxy spectra to template star spectra. The fitting methodology is
explained in detail elsewhere (see,
e.g., {van Dokkum} \& {Franx} 1996; {Kelson} {et~al.} 2000b; {van Dokkum} \& {Stanford} 2003). Briefly,
the template stars were
broadened such that their spectral resolution matches the instrumental
resolution of the observations. Residuals of night sky lines, Balmer
absorption lines and (in the case of \clusa--2014) emission lines were
masked in the fits. The wavelength region used in the fit was
4100\,\AA\,--\,4650\,\AA; changing the fitting region has a negligible effect
on the measured dispersions. For consistency with previous work
(e.g., {van Dokkum} \& {Franx} 1996) a K0 giant star was adopted as the
template star. Varying the
template star and the continuum filtering produces $\lesssim 5$\,\%
variations in the dispersions.

For each galaxy (except \clusa--47) velocity dispersions were measured
from the unweighted average spectrum of the five central rows,
corresponding to a $1\farcs 1 \times 1\farcs 1$ square aperture.
For consistency with previous work the measured dispersions were
corrected to the equivalent of a $3\farcs 4$ diameter
aperture at the distance of the Coma cluster, using the empirical
logarithmic correction given by {J{\o}rgensen}, {Franx}, \&  {Kj\ae{}rgaard} (1995b). For $z\approx 0.5$
the correction is approximately $7$\,\%. Corrected dispersions are
listed in Table 2. The median random error is 8\,\%, and
in every case the error is less than 15\,\%.

\subsection{Spatially Resolved Kinematics}
\label{kinr.sec}

In vdMvD06a we present dynamical models of the
galaxies, derived from spatially-resolved kinematics and
photometry. Among other parameters,
these models provide $M/L$ ratios
of the galaxies free of many of the assumptions that enter
the FP analysis. The radial profiles of velocity and
velocity dispersion were determined in the
following way.
Velocity dispersion measurements were made at 7 positions
along the slit: the average of rows $-7,-6,-5$; the average
of rows $-4,-3,-2$; row $-1$; the central row 0;
row +1; the average of rows
$2,3,4$; and the average of rows $5,6,7$. As the pixel size
is $0\farcs 215$ these positions
correspond to $-1\farcs 29$, $-0\farcs 64$, $-0\farcs 22$,
$0\arcsec$, $0\farcs 22$, $0\farcs 64$, and $1\farcs 29$
from the center of the galaxy. Radial
velocities were measured  using {\sc xcsao},
as in some cases velocities can still be measured when the S/N
is too low to measure dispersions. Dispersions were deemed
unreliable when the random error exceeds $20$\,\%; velocities
were discarded when the {Tonry} \& {Davis} (1979) R-value is
lower than 3. The resulting profiles are shown in vdMvD06a.

\begin{figure*}[t]
\epsfxsize=16.5cm
\epsffile[-10 90 551 671]{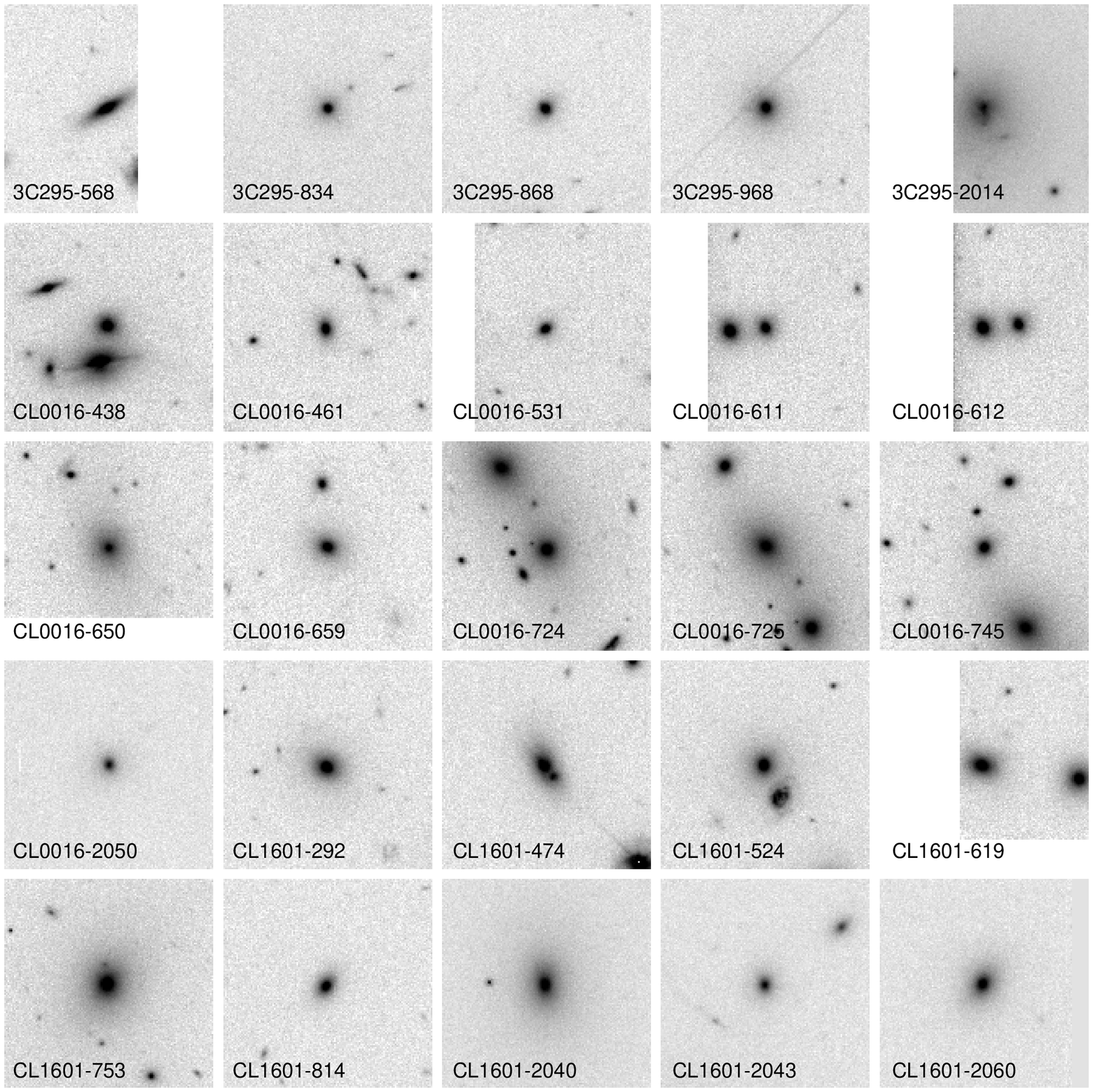}
\caption{\small
{\em Hubble Space Telescope} WFPC2 images of the 25 cluster early-type
galaxies observed spectroscopically.
\label{stamps1.plot}}
\end{figure*}

\section{Photometry}

\subsection{WFPC2 Images}

The three clusters were observed with WFPC2 on {\em HST} as part of
the MORPHS cluster program. Details of the observations and data
characteristics are given in {Smail} {et~al.} (1997). \clusa\ and \clusc\
were observed in one band ($R_{702}$) only; \clusb\ was observed in
two filters ($V_{555}$ and $I_{814}$). Total exposure times are listed
in Table 1.

We obtained the raw data from the {\em HST} archive. For each cluster,
the exposure time was divided over two positions offset by
$\approx 2\arcsec$. The shifts are not integer numbers of pixels, and
rather than combining the two positions we reduced each pointing
separately. This procedure has the advantages that no interpolation is
required, and that we can assess the uncertainties in derived
parameters from two independent datasets, with different sub-pixel
sampling. The IRAF {\sc crrej} task was used to combine individual
frames at each position. Remaining cosmic rays and hot pixels were
removed using {\sc L.A.Cosmic}.

Images of the 25 cluster galaxies are shown in Fig.\ \ref{stamps1.plot}.
As expected, most of the galaxies have very regular morphologies with
no obvious spiral arms or other finestructure. There are
three notable exceptions: \clusa--568 is the S0/Sb galaxy
added to test our ability to measure rotation curves; \clusa--2014
(the central cluster galaxy) has structure in its core, very likely
associated with its active nucleus; and \cluscs--474 consists of
two overlapping galaxies. Both components of \cluscs--474 show
morphological disturbances, strongly suggesting this is a merger
in progress. The object is flagged as ``interaction/merger?'' in
the {Smail} {et~al.} (1997) catalog. Interestingly the galaxy has an
early-type spectrum, similar to the red mergers in the
$z=0.83$ cluster MS\,1054--03 ({Tran} {et~al.} 2005a).


\subsection{Structural Parameters}
\subsubsection{Fitting}
\label{strucfit.sec}

The galaxy images were fitted with two-dimensional models, created by
convolving $r^{1/4}$-law light distributions with point spread
functions (PSFs). A separate PSF was used for each object, as the
WFPC2 PSF depends on the position of the object on the chip. The PSFs
were generated with the Tiny Tim 6.1a software ({Krist} 1995).  The
fitting procedure is described in {van Dokkum} \& {Franx} (1996). Fit parameters in
the $\chi^2$ minimalization are the $x,y$ position, the effective
radius $r_e$, the surface brightness at the effective radius $I_e$,
the position angle, the ellipticity, and the sky value. Nearby
objects and the edges of ccds were masked in the fits.
We inspected the residual images, created by subtracting the best
fitting models from
the data, to assess the quality of the
fit. In most cases the
model provides a good fit to the data; exceptions
are the S0/Sb galaxy \clusa--568
and the active nucleus \clusa--2014.

We empirically determined the errors in the fitted parameters by
comparing the values derived from the two independent pointings, as
the formal errors from the $\chi^2$ fit
do not include the effects of undersampling.  The
results are shown in Fig.\ \ref{compstruc.plot}. The uncertainties in
$r_e$ and $I_e$ are highly correlated, and as is well known the
correlation is almost parallel to the relation $I_e \propto
r_e^{-1.2}$ (solid line) which enters the FP
(see, e.g., J{\o}rgensen, {Franx}, \&  {Kj\ae{}rgaard} 1995a; {van Dokkum} \& {Franx} 1996; {Kelson} {et~al.} 2000a).
Not surprisingly the correlation is also well described by the expected
relation for total flux conservation ($I_e \propto r_e^{-2}$;
broken line). The rms uncertainty in a single measurement is
$0.012$ in $\log r_e$, $0.018$ in $\log I_e$, and
$0.003$ in the FP combination $\log r_e + 0.8 \log I_e$.
These uncertainties are negligible compared to the random
errors in the velocity dispersions, and also to the
systematic errors in the photometric calibration (see \S\,\ref{calib.sec}).

\vbox{
\begin{center}
\leavevmode
\hbox{%
\epsfxsize=8.5cm
\epsffile{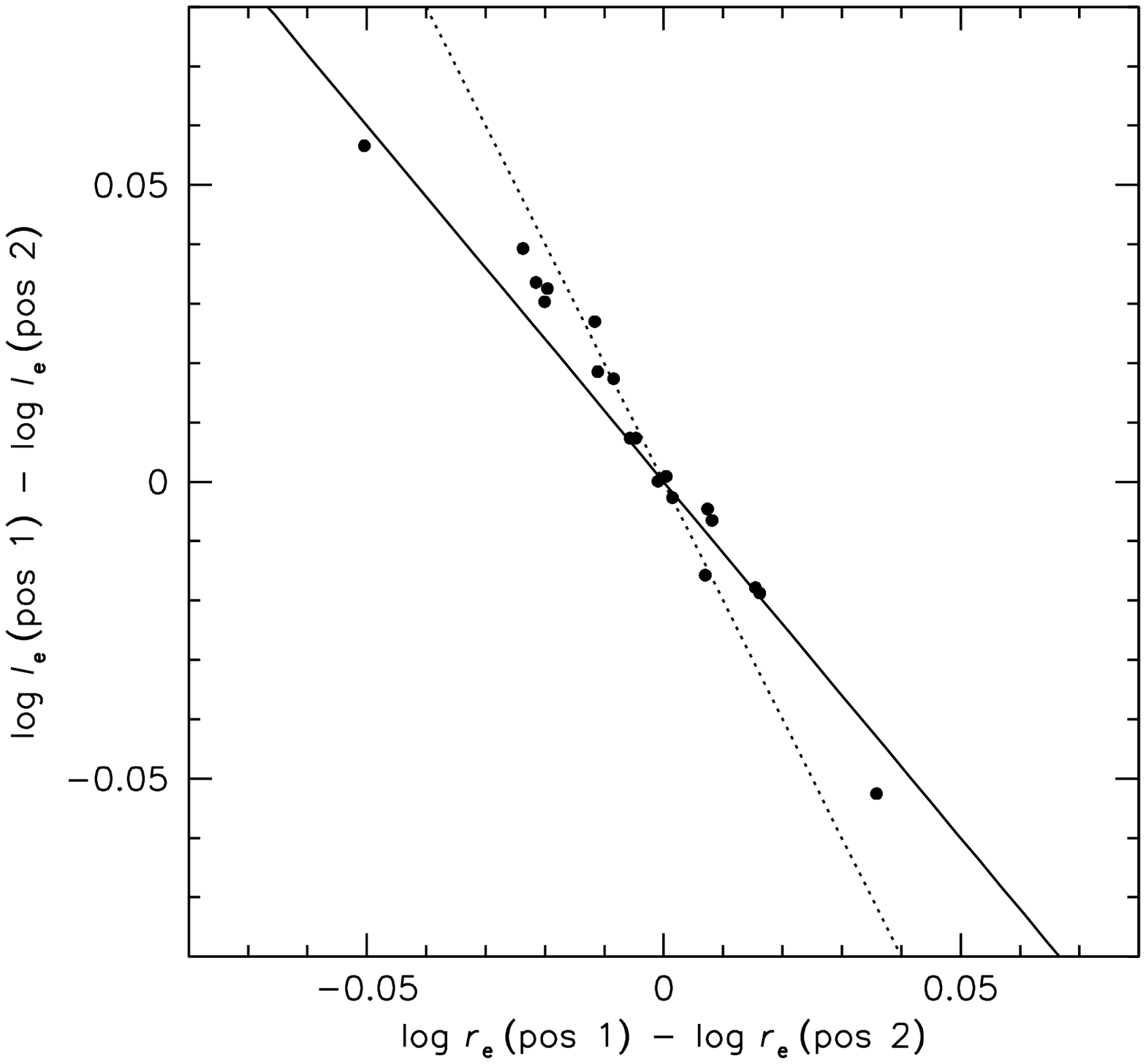}}
\figcaption{\small
Comparison of effective radii $r_e$ and surface brightnesses $I_e$
measured from two independent {\em HST} pointings. The errors
in the two parameters are highly correlated. The broken line
is the relation $I_e \propto r_e^{-2}$, and the solid line
is the relation $I_e \propto r_e^{-1.2}$ which enters the fundamental plane.
\label{compstruc.plot}}
\end{center}}

\subsubsection{Calibration}
\label{calib.sec}

Measured effective radii were converted to arcseconds using a pixel
scale of $0\farcs 09993$\,pixel$^{-1}$ for the Wide Field camera, and
$0\farcs 04555$\,pixels$^{-1}$ for the Planetary Camera. They were
converted to kpc using our adopted cosmology. We note that the
analysis in \S\,\ref{mlevo.sec}
depends on the value of $\Omega$ but not on the Hubble constant
(see, e.g., {van Dokkum} {et~al.} 1998a).

Surface brightnesses were converted to the {\em HST} Vega system
using the appropriate zeropoints for each of the WFPC2 chips, as obtained
from the WFPC2 Data Handbook,\footnote{See
http://www.stsci.edu/instruments/wfpc2/Wfpc2\_dhb/WFPC2\_longdhbcover.html.}
and corrected for Galactic extinction using the {Schlegel}, {Finkbeiner}, \&  {Davis} (1998) maps.
The extinction is highest in the \clusb\ field ($\approx 0.11$\,mag
in the $I_{814}$ band). We did not correct for the long vs.\ short
anomaly, as our fields are not very crowded, or the position-dependent
charge transfer efficiency (CTE), as the backgrounds are relatively high.

For a meaningful comparison of galaxies observed at different redshifts
the photometry has to be transformed to a common rest-frame band. At
$z\sim 0.5$ the observed $R_{702}$ band is close to the rest-frame
$B$ band (denoted $B_z$).
Following the methodology of {van Dokkum} \& {Franx} (1996)
we derived the following transformations from observed to rest-frame
magnitudes:
\begin{eqnarray}
\label{trafo.eq}
B_z & = & R_{702} + 0.23 (V_{555} - R_{702}) + 0.74 \\
B_z & = & I_{814} + 0.30 (V_{555} - I_{814}) + 0.89 \\
B_z & = & R_{702} + 0.05 (V_{555} - R_{702}) + 0.85,
\end{eqnarray}
for \clusa, \clusb, and \clusc\ respectively.
For the synthetic $B$ band we used the
Bessell (1990) $BX$ filter. This
filter is very similar to
the Buser \& Kurucz (1978) $B_2$ filter, which is
used by Bruzual \& Charlot (2003) for their stellar population
synthesis calculations.

Using templates
from {Coleman}, {Wu}, \& {Weedman} (1980) we find that the transformations
are independent of spectral type to $\approx 0.03$.
However, the
colors that enter the transformations are obviously a strong function
of spectral type. Measured colors are only available
for \clusb\ ({Smail} {et~al.} 1997), as the other clusters were observed in only
one filter. For \clusa\ and \clusc\
we used synthetic colors derived from a redshifted
{Coleman} {et~al.} (1980)
E/S0 template, as the measured colors for \clusb\ are a good match to this
template. We find $V_{555}-R_{702}=
1.71$ for \clusa, and $V_{555}-R_{702}=1.81$ for \clusc.

The uncertainties in the calibration are a combination of
uncertainties in the WFPC2 zeropoints ($\approx 0.03$\,mag),
the long-short anomaly and CTE effects ($\approx 0.04$\,mag),
the extinction ($\approx 0.02$\,mag), and the conversion to
rest-frame $B$ magnitudes. The uncertainty
introduced by the lack of measured colors is only a few percent
for \clusc, as the color term is small. However, the effect can
be substantial
for the lower redshift cluster \clusa:
using an Sb/c template rather than an E/S0 template
gives $B_z$ magnitudes that are $\approx 0.15$\,mag brighter.
Hence we estimate that the typical uncertainty $\approx 0.10$
mag for this cluster. We assume that the systematic errors
are uncorrelated, and
can be added in quadrature (see, e.g., Freedman et al.\
2001). The combined calibration
errors in the surface brightnesses
are $\approx 0.11$\,mag for \clusa\ and
$\approx 0.06$\,mag for \clusb\ and \clusc.


\begin{figure*}[t]
\epsfxsize=17.5cm
\epsffile[35 472 513 637]{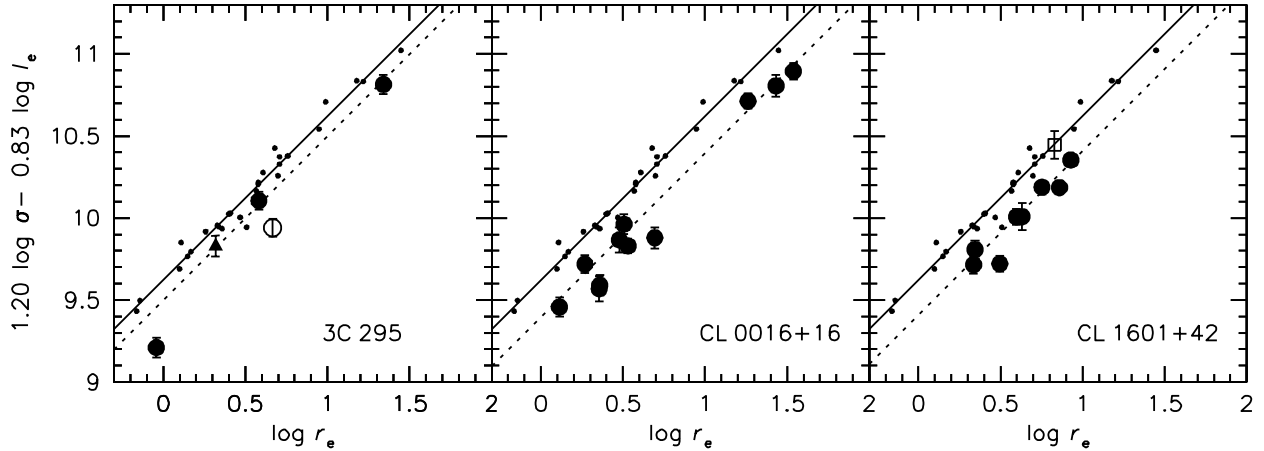}
\caption{\small
Edge-on projection of the FP in the three clusters. Filled circles are
E and E/S0 galaxies, the filled triangle is the S0 galaxy \clusa--868,
the open circle is the S0/Sb galaxy \clusa--568, and the open square
is the merger \cluscs--474. Dots are galaxies
in the nearby Coma cluster. The solid line is a fit to Coma, and
the broken lines show the best fitting offsets from the Coma
FP for each cluster.
\label{fp.plot}}
\end{figure*}

\section{The Fundamental Plane}

The Fundamental Plane 
is an
empirical relation between the effective radii $r_e$, surface brightness
at the effective radii $I_e$, and the velocity dispersions $\sigma$ of 
early-type galaxies of the form
\begin{equation}
\log r_e = a \log \sigma + b \log I_e + c
\end{equation}
({Djorgovski} \& {Davis} 1987). The coefficients $a$ and $b$ depend on
wavelength, sample selection, and fitting
method (see, e.g., {Bernardi} {et~al.} 2003a).
{J{\o}rgensen}, {Franx}, \&  {Kj\ae{}rgaard} (1996) find $a=1.20 \pm 0.06$ and $b=-0.83 \pm 0.02$
in the $B$ band. We define $\log I_e \equiv \mu_e/-2.5$,
with $\mu_e$ the rest-frame
$B$ band surface brightness at the effective radius
in mag\,arcsec$^{-2}$, corrected for $(1+z)^4$ cosmological
surface brightness dimming. We note that for an r$^{1/4}$-law fit the
relation between $\mu_e$ and $\langle \mu\rangle_e$, the average
surface brightness within the effective radius, is given by
$\mu_e = \langle \mu \rangle_e + 1.393$.

\subsection{Edge-on Projection}

The edge-on projections of the FPs in the three clusters are shown
in Fig.\ \ref{fp.plot}. Solid circles are E and E/S0 galaxies,
the solid triangle is an S0 galaxy, the open square is a merger
system, and the open circle is the
S0/Sb galaxy that was added to test whether we can measure
rotation curves. Small dots are galaxies in the nearby
Coma cluster from {J{\o}rgensen} {et~al.} (1996) (see \S\,\ref{coma.sec}).

The FP is well defined in each cluster: the elliptical galaxies
in the $z\approx 0.5$ clusters qualitatively follow very similar relations
as the Coma galaxies.
The FPs are offset with respect to Coma because of passive
luminosity evolution of the stellar populations in the galaxies
(see, e.g., van Dokkum \& Franx 1996).
The S0/Sb galaxy \clusa--568 and the merger \cluscs--474 are outliers.
The spiral galaxy falls below the relation defined by the other galaxies
in \clusa,
consistent with results for spiral galaxies in the $z=0.33$ cluster
CL\,1358+62 ({Kelson} {et~al.} 2000c). The merging galaxy falls above the
\clusc\ relation, possibly because its velocity dispersion
is overestimated due to the presence of light
from its companion in the slit. Both galaxies were excluded from the
analysis in \S\,\ref{mlevo.sec}.

\begin{figure*}[t]
\epsfxsize=17.5cm
\epsffile[32 469 510 637]{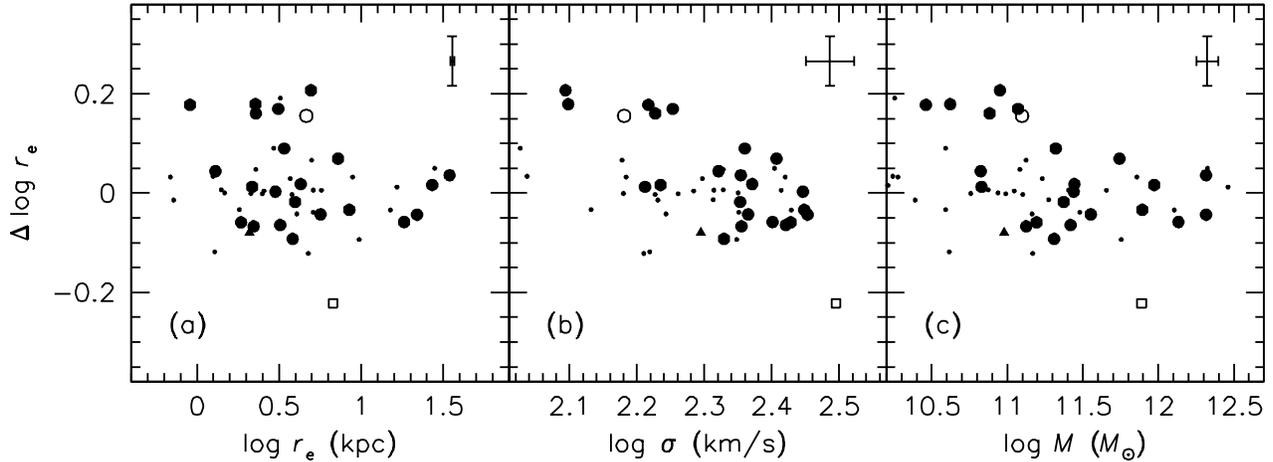}
\caption{\small
Correlation of the residual from the FP with effective radius
(a), velocity dispersion (b), and mass (c), for galaxies in
the three clusters. Symbols have the same
meaning as in Fig.\ \ref{fp.plot}.
Galaxies with low masses or low velocity dispersions
appear to systematically deviate with respect to the Coma cluster
sample.
\label{scatter.plot}}
\end{figure*}

\subsection{Scatter and Tilt of the Fundamental Plane}
\label{scatter.sec}

The galaxy samples in the individual clusters are too small to
reliably determine the scatter in the FP. Hence the data for
the three clusters are combined. For each galaxy the
residual from the FP was calculated using
\begin{equation}
\Delta_{\rm FP}\,(\log r_e) = \log r_e - (1.20 \log \sigma  - 0.83 \log I_e + c)
 - 0.461 (z-0.024).
\end{equation}
This procedure assumes that the
intrinsic cluster-to-cluster scatter in the zeropoint of the FP
is small compared to
the galaxy-to-galaxy scatter within the clusters.
The redshift term compensates for differential luminosity evolution
over the redshift range spanned by the clusters, and the zeropoint
$c = -9.626$ is that of the $z=0.024$ Coma cluster (see \S\,\ref{mlevo.sec}).

The scatter is calculated using the biweight estimator and its
uncertainty is determined from bootstrap resampling.
For the full sample of E and E/S0 galaxies (excluding the merger
\cluscs--474) the scatter is $0.095 \pm 0.028$ in
$\log r_e$.
This scatter is slightly higher than that measured
in local clusters; e.g., {J{\o}rgensen} {et~al.} (1996) find an rms scatter of
$0.071$ in $\log r_e$. We use Monte Carlo simulations
to determine whether the measured
scatter can be fully explained by measurement errors
(which are dominated by the errors in $\log \sigma$).
The errors imply an expected scatter of $0.048 \pm 0.009$,
and we conclude that the intrinsic scatter in this sample
is $0.082 \pm 0.028$ in $\log r_e$.

We investigate the cause of this scatter in Fig.\ \ref{scatter.plot},
which shows the relation of the residuals from the FP with
effective radius, velocity dispersion, and mass. Masses
were determined using
\begin{equation}
\label{mass.eq}
\log M = 2 \log \sigma + \log r_e + 6.07.
\end{equation}
The masses are approximate, as the constant in Eq.\ \ref{mass.eq}
depends on the details of the galaxy model, such as the brightness
profile and the internal dynamical structure.
There are clear systematic trends in the $z\approx 0.5$ sample
which are not present in the Coma sample. The residuals
correlate with velocity dispersion and with mass: a Spearman
rank test gives a probability of $99.8$\,\% that
$\Delta \log r_e$ and $\sigma$ are correlated, and a
probability of $96$\,\% that $\Delta \log r_e$ and $M$ are
correlated. Correlations with
$\log r_e$ and $\log I_e$ are not significant according to the
Spearman rank test.
When the sample is limited to galaxies
having $\sigma>200$\,\kms\ the observed scatter in $\Delta
\log r_e$ reduces to $0.055 \pm 0.019$; galaxies with
$M>10^{11}$\,\msun\ give a scatter of $0.064 \pm 0.020$. Both
values are consistent with the scatter due to measurement
errors alone, and we conclude that the intrinsic scatter in
the full sample of E/S0 galaxies is caused by the
galaxies with the lowest masses.

Similar correlations were found by {Wuyts} {et~al.} (2004) for
a cluster at $z=0.58$ and by
{J{\o}rgensen} {et~al.} (2006) for two clusters
at $z=0.83$ and $z=0.89$.
Furthermore, recent studies of field early-type
galaxies at $0.5<z<1.1$ have found the same qualitative trends
({Treu} {et~al.} 2005b; {van der Wel} {et~al.} 2005; {di Serego Alighieri} {et~al.} 2005). The usual interpretation
of these trends is that they are due to a combination of
selection effects and evolution in the ``tilt'' of the FP
(see, e.g., {Treu} {et~al.} 2005a). The evolution in the tilt
is taken as evidence for differential evolution in the
$M/L$ ratio for galaxies of different mass, resulting from
an underlying age-mass relation (with low mass early-type
galaxies being younger than high mass early-type galaxies --
see, e.g., {Treu} {et~al.} 2005a; {van der Wel} {et~al.} 2005).
In vdMvD06b we show that these observed trends may be
in part
due to structural or kinematic evolution of low
mass galaxies, as opposed to evolution of their $M/L$ ratios.
In the following
we restrict the discussion to galaxies with
masses $>10^{11}$\,\msun. As shown in vdMvD06b
FP evolution is an unbiased measure of $M/L$ evolution
for galaxies in this mass range.

Finally, we note that {Kelson} {et~al.} (2000c) and {Moran} {et~al.} (2005)
do not find evidence for a change in the tilt of the FP for
the clusters CL\.1358+62 at $z=0.33$ and CL\,0024+16 at
$z=0.39$ respectively, despite the use of large, high quality datasets.
We also note that {Wuyts} {et~al.} (2004) and {Moran} {et~al.} (2005) report an
increased intrinsic scatter in the FP compared to nearby
clusters even for high mass galaxies,
which is not seen in our sample or in
CL\,1358+62 ({Kelson} {et~al.} 2000c).
The available evidence suggests that the FPs of intermediate redshift
clusters show a range of properties when studied in detail,
perhaps due to the stochastic nature of the growth of clusters
(see, e.g., {Tran} {et~al.} 2005b; {Moran} {et~al.} 2005).
Studies of individual clusters should therefore
be interpreted with caution.

\section{Evolution of the Mean M/L Ratio}
\label{mlevo.sec}

\subsection{Procedure}
\label{procedure.sec}

Evolution of the zeropoint
of the FP can be interpreted as a systematic change of
the mean $M/L$ ratio with redshift.
The conversion of zeropoint offsets to offsets in
$M/L$ ratio assumes that early-type galaxies form a homologous
family and that the FP is a manifestation of an underlying relation
between the $M/L$ ratio of galaxies and other parameters
(e.g., Faber et al.\ 1987). Starting from
the empirical FP relation and assuming
$M \propto \sigma^2 r_e$ and $L \propto I_e r_e^2$, this underlying
relation is
\begin{equation}
M/L \propto \sigma^{2+a/b}r_e^{-(1+b)/b}
\end{equation}
(e.g., {Kelson} {et~al.} 2000c). The {J{\o}rgensen} {et~al.} (1996) $B$-band
coefficients imply that $M/L$ is largely a function of
mass: $M/L \propto M^{0.28} r_e^{-0.08}$.
Different coefficients imply different relations; e.g.,
using a large sample of galaxies
drawn from the {\em Sloan Digital Sky Survey} (SDSS)
Bernardi et al.\ (2003a) find that $M/L$ correlates with effective
radius with little or no dependence on $\sigma$.

Rather than determine relations between $M/L$ and other
observables we follow the usual practice of determining
the evolution in $M/L$ directly from the fundamental plane
offset with respect to $z\approx 0$:
\begin{equation}
\label{dml.eq}
\Delta \log (M/L)_z = \log (M/L)_z - \log (M/L)_0 = (c_z - c_0)/b.
\end{equation}
This procedure assumes that the observed change in intercept of the
FP is caused by evolution of the $M/L$ ratio and that the
coefficients $a$ and $b$
do not depend on the redshift. In vdMvD06b it is demonstrated
that FP evolution thus defined yields an unbiased measure
of $M/L$ evolution for galaxies with velocity dispersions
$\gtrsim  200$\,\kms\ or masses $\gtrsim 10^{11}$\,\msun.
In practice we use the Coma cluster of galaxies to define the
zeropoint $c_0$ (see Appendix \ref{lit.sec}). This choice is arbitrary,
as adding a constant to all value of $\Delta \log (M/L)_z$
does not change the results of our subsequent analysis in any way.

The coefficient $c_z$ for each cluster is determined by
calculating the residual from the FP 
for each early-type galaxy with $M>10^{11}$\,\msun:
\begin{equation}
\label{czi.eq}
c_{z,i} = \log r_e - (a \log \sigma + b \log I_e).
\end{equation}
The cluster offset $c_z$ is
the biweight mean ({Beers}, {Flynn}, \& {Gebhardt} 1990) of the distribution of
$c_{z,i}$.
For consistency with {J{\o}rgensen}
{et~al.} (1996) we use $a=1.20$ and $b=-0.83$
in Eqs.\ \ref{dml.eq} and \ref{czi.eq}.
As we show later the results are only very weakly dependent on the
values of these parameters. For each cluster a random error and
a systematic error is determined. The random error is the
formal uncertainty in the biweight mean, or the expected
uncertainty due to the errors in the individual measurements,
if the latter exceeds the former. As discussed in \S\,\ref{strucfit.sec}
random errors in the velocity dispersions dominate over those
in the structural parameters. The minimum random error in
$c_z$ is therefore approximately
$\overline{\delta \sigma}/\sqrt{n}$, with $\overline{\delta \sigma}$ the
average random uncertainty in the velocity dispersion and $n$ the
number of galaxies in the sample. Systematic errors are described
for each cluster individually below. 

\subsection{Other Clusters}
\label{otherc.sec}

Including the three clusters presented here
there are now more than a dozen clusters in the redshift range
0.2 to 1.3 for which FP measurements have been made.
Furthermore,
at low redshift the SDSS has provided important new information
unhampered by uncertainties in peculiar motions and small number
statistics. We combine published data with our own to compile 
a sample consisting of the Coma cluster, the SDSS sample, and
fourteen distant clusters with homogeneous FP measurements.
With one exception,
the criteria for including distant clusters are 1) published
data for individual galaxies, and 2) sufficient information to
bring the data to a consistent system (i.e., colors,
aperture corrections, etc). The exception is
RX\,J1226+33 at $z=0.892$: data of individual
galaxies in that cluster are not yet published, but they
were provided to us by I.\ J\o{}rgensen.
Details on the individual clusters and the
derivation of systematic uncertainties are given in Appendix
\ref{lit.sec},
and the offsets are listed in Table 3.
The uncertainties include neither the propagation of
uncertainties
in $\Omega_m$ and $\Omega_{\Lambda}$, nor the propagation of
uncertainties in distance due to cosmic variance in $H_0$
(i.e., the expansion factor having somewhat different value
when averaged over small scales than when averaged over large
scales). This simplification is justified by the fact that
these uncertainties are generally not the dominant uncertainty
in the analysis (see vdMvD06b).

\begin{small}
\begin{center}
{ {\sc TABLE 3} \\
\sc $M/L_B$ Offsets} \\
\vspace{0.1cm}
\begin{tabular}{lcccc}
\hline
\hline
Sample & $z$ & $N$ & $\Delta \log (M/L_B)$ & $\pm$ \\
\hline
Coma       & 0.024 & 16  &   0.000 & 0.029 \\
SDSS, $n_{\rm den}>20$   & 0.109 & 171 & $+0.021$ & 0.031 \\
Abell 2218 & 0.176 & 8   & $+0.009$ & 0.037 \\
Abell 665  & 0.183 & 5   & $-0.006$ & 0.040 \\
Abell 2390 & 0.228 & 5   & $-0.035$ & 0.065 \\
CL\,1358+62 & 0.327 & 16 & $-0.162$ & 0.029 \\
CL\,0024+16 & 0.391 & 6  & $-0.160$ & 0.040 \\
3C\,295     & 0.456 & 2  & $-0.158$ & 0.053 \\
CL\,1601+42 & 0.539 & 8  & $-0.321$ & 0.043 \\
CL\,0016+16 & 0.546 & 7  & $-0.281$ & 0.032 \\
MS\,2053--04 & 0.583 & 8 & $-0.287$ & 0.056 \\
MS\,1054--03 & 0.831 &12 & $-0.427$ & 0.040 \\
RX\,J0152--13 & 0.837 &13& $-0.449$ & 0.048 \\
RX\,J1226+33 & 0.892 & 5 & $-0.558$ & 0.056 \\
RDCS\,1252--29 & 1.237 & 4&$-0.586$ & 0.116 \\
RDCS\,0848+44 & 1.276 & 2 & $-0.542$ & 0.069 \\
\hline
SDSS, $n_{\rm den}<10$ & 0.114 & 1519 & $-0.010$ & 0.031 \\
Field  & 0.380 & 8  & $-0.164$ & 0.058 \\
--- &    0.468 & 10 & $-0.258$ & 0.061 \\
--- &    0.563 & 10 & $-0.312$ & 0.071 \\
--- &    0.747 & 10 & $-0.445$ & 0.064 \\
--- &    0.563 & 10 & $-0.312$ & 0.071 \\
--- &    0.747 & 10 & $-0.445$ & 0.064 \\
--- &   0.844  & 10 & $-0.473$ & 0.077 \\
--- &  0.851 & 10 & $-0.538$ & 0.053 \\
--- &  0.951 & 10 & $-0.476$ & 0.105 \\
--- & 1.016 & 10 & $-0.626$ & 0.052 \\
--- & 1.110 & 9 & $-0.598$ & 0.055 \\
\hline
\end{tabular}
\end{center}
\end{small}

\subsection{Observed Evolution and Scatter}
\label{evosub.sec}

The observed evolution of the $M/L_B$ ratio of massive
cluster galaxies is shown in Fig.\ \ref{mlz.plot}.
There is a clear relation, with galaxies in
clusters at high redshift having lower $M/L_B$ ratios than those
at low redshift, consistent with previous studies of
smaller samples (e.g., {van Dokkum} \& {Franx} 1996; {van Dokkum} {et~al.} 1998a; {Kelson} {et~al.} 1997; {Holden} {et~al.} 2005). The best-fitting linear function has a slope
\begin{equation}
d \log (M/L_B)/dz =-0.555 \pm 0.042,
\end{equation}
and is indicated in Fig.\ \ref{mlz.plot}.
This evolution is stronger than inferred from previous studies that
were based on smaller samples. For example, van Dokkum \& Stanford
(2003) find a slope of $-0.460 \pm 0.039$ and 
Holden et al.\ (2005) find $-0.426 \pm 0.026$. This difference is
largely due to
sample size: if we limit the sample to the six clusters from
van Dokkum \& Stanford or the
seven clusters from Holden et al., we find slopes of
$-0.454 \pm 0.052$ and $-0.463 \pm 0.056$ respectively, consistent with
their results. We also note that our treatment of systematic errors
is different from previous studies, which leads to slightly different
fits (with different quoted uncertainties), even for the same data.

\vbox{
\begin{center}
\leavevmode
\hbox{%
\epsfxsize=8.5cm
\epsffile{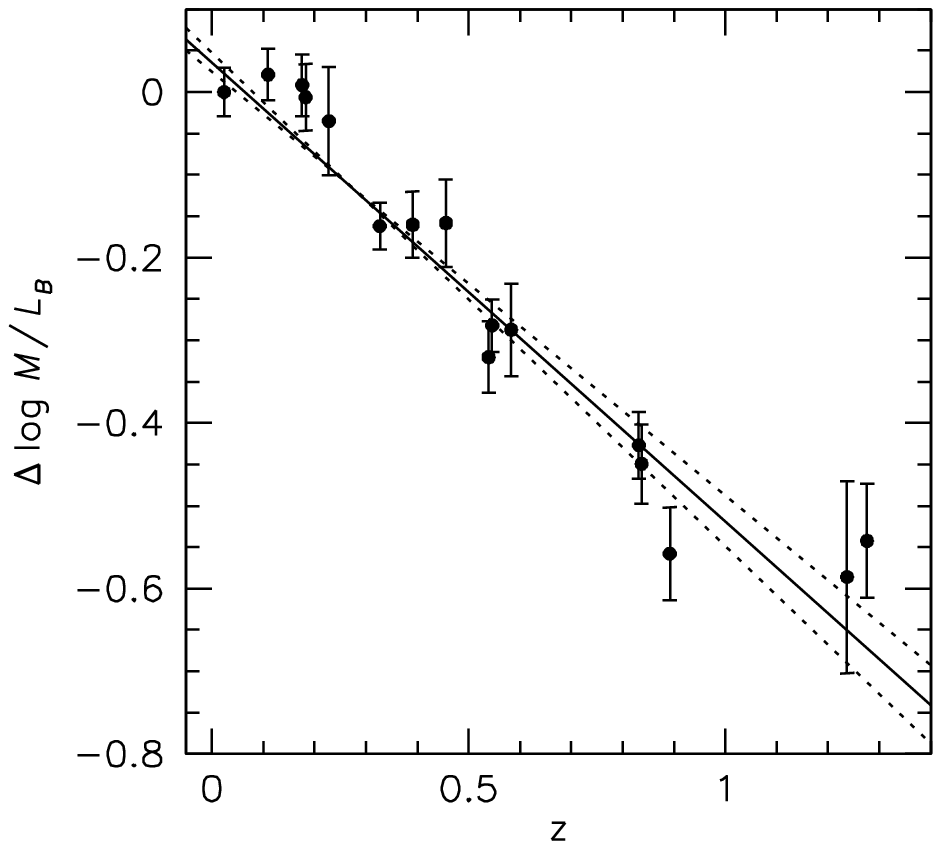}}
\figcaption{\small
Redshift evolution of the mean $M/L_B$ ratio of cluster galaxies with masses
$M>10^{11}\,M_{\odot}$ in our composite sample.
The solid line shows the best fitting linear
function, with slope
$d\log (M/L_B)/dz= -0.555$. Dotted lines
indicate the $\pm 1 \sigma$ uncertainty in the slope of
the relation.
\label{mlz.plot}}
\end{center}}

The quoted uncertainty in the fit does not
include all sources of systematic error. We first investigate the
effects of the choice of coefficients of the FP
(which have some uncertainty; see, e.g., {Bernardi} {et~al.} 2003a; {Cappellari} {et~al.} 2006). Varying
$a$ from $1.0$ to $1.5$ produces  best fitting slopes ranging
from $-0.564$ to $-0.542$ and varying $b$ from $-1.1$ to
$-0.6$ gives slopes in the range $-0.539$ to $-0.548$.
We conclude
that the precise FP coefficients do not have a large effect on the derived
redshift evolution.

Another source of uncertainty is the selection
of the clusters. Some are  selected in the optical (e.g., CL\,1601+42
and CL\,0016+16), others in X-rays (e.g., MS\,2053--04 and
MS\,1043--03). Furthermore, the low redshift Coma cluster and
Bernardi et al.\ sample may not be representative for the
descendants of the (very massive) high redshift clusters in the sample.
The data are insufficient to investigate the evolution of
subsamples selected by X-ray luminosity, mass, or other parameters.
However, we can assess the effects of removing datapoints,
and determine whether the fit is driven by some
individual clusters. Removing the Bernardi sample changes the
best fitting slope by $+0.009$ to $-0.546$. Removing Coma
changes the slope by $-0.014$. Removing both Coma and the Bernardi
sample changes the slope by $-0.005$. The latter test effectively
transfers the low redshift comparison point to the very
rich $z\approx 0.2$ clusters Abell 665, Abell 2218, and Abell 2390,
which may be appropriate for the descendants of massive clusters
like MS\,1054--03 at $z=0.83$. All these changes are very small,
and well within the $1\sigma$ formal uncertainty in the fit.
The cluster with the largest effective weight in the fit is
RDCS\,0848+44 at $z=1.276$: removing this cluster changes the
slope by $-0.052$ to $-0.607$. The large influence of
this single cluster highlights
the importances of obtaining more FP measurements
for cluster galaxies at $z>1$.

The sample is sufficiently large to determine the scatter in
the best-fitting linear relation. The $\chi^2$ of the fit is
20.4 with 14 degrees of freedom, which implies that a linear
function is an adequate description of the data and that the
scatter can be explained by the measurement uncertainties.
The maximum allowed intrinsic cluster-to-cluster scatter can be
determined by requiring that $\chi^2>7.8$, which for 14
degrees of freedom corresponds to a one-sided probability
$>0.1$. Iteratively adding uncertainty due to intrinsic scatter in
quadrature to the errors listed in Table 3 we find
that $\chi^2<7.8$ for an additional uncertainty of $>0.057$.
The 90\,\% confidence upper limit on the intrinsic
cluster-to-cluster scatter is therefore $0.057$ in
$\log (M/L_B)$.

\section{Implications}

\subsection{The Star Formation Epoch of Massive Cluster Galaxies}

The observed evolution of the $M/L_B$ ratio of early-type
galaxies depends on cosmological parameters, the
star formation histories of the galaxies, their initial mass
function (IMF) and metallicity, possible changes in dust content
with redshift, and selection effects.
Although the dependence on cosmological parameters can, in principle,
be exploited (see, e.g., {Pahre}, {Djorgovski}, \& {de  Carvalho} 1996; {van Dokkum} {et~al.} 1998a; {J{\o}rgensen} {et~al.} 1999; {Lubin} \& {Sandage} 2001),
we assume here that
the evolution of galaxies and their stellar populations is currently
more uncertain than the values of $\Omega_m$ and $\Omega_{\Lambda}$
(which are the relevant parameters). We also assume that dust
is either absent in early-type cluster galaxies or non-evolving
over the redshift range $0<z<1.3$.
The star formation history is parameterized with a single age for
the entire stellar population. This is almost certainly incorrect
as most early-type galaxies likely have extended and complex star
formation histories at early times (e.g., {Nagamine} {et~al.} 2005; {Knudsen} {et~al.} 2005; {Papovich} {et~al.} 2006). However, as shown by, e.g., {van Dokkum} {et~al.} (1998b)
the evolution of a complex stellar population can be
well approximated by that of a single-age stellar population of the
same luminosity-weighted age, provided that star formation has
ceased $\gtrsim 1$\,Gyr before the epoch of observation.

Figure \ref{modfit.plot} shows the predicted age-dependence of the
$M/L_B$ ratio for different assumptions about the IMF, the metallicity,
and the implementation of horizontal branch stars in the population
synthesis code. Solid lines are various {Bruzual} \& {Charlot} (2003) models, using the
{Salpeter} (1955) and {Chabrier} (2003) IMFs and [Fe/H] ranging
from Solar to $+0.56$. Broken lines are various models of
{Maraston} (2005), whose implementation of the late
stages of stellar evolution differs from that of {Bruzual} \& {Charlot} (2003).
The Maraston models that are shown are for {Salpeter} (1955)
and {Kroupa} (2001) IMFs, metallicities ranging from
Solar to $+0.67$, and four different implementations of
the horizontal branch. All models were normalized in order to emphasize
the predicted rate of evolution rather than the absolute $M/L$ ratios
at any given time (which vary greatly between the models).

The predictions are very similar, demonstrating that the predicted
evolution is not very dependent on the metallicity, IMF, or
particulars of the stellar population synthesis code. An important
caveat is that the {Salpeter} (1955), {Kroupa} (2001), and
{Chabrier} (2003) IMFs, while different, were all derived from
Galactic data, and the IMF of the (progenitors of) early-type
galaxies may have been
weighted more toward very massive stars (see, e.g., {Larson} 1998).
We will return to this issue in \S\,\ref{conclusions.sec}.

\vbox{
\begin{center}
\leavevmode
\hbox{%
\epsfxsize=8.5cm
\epsffile{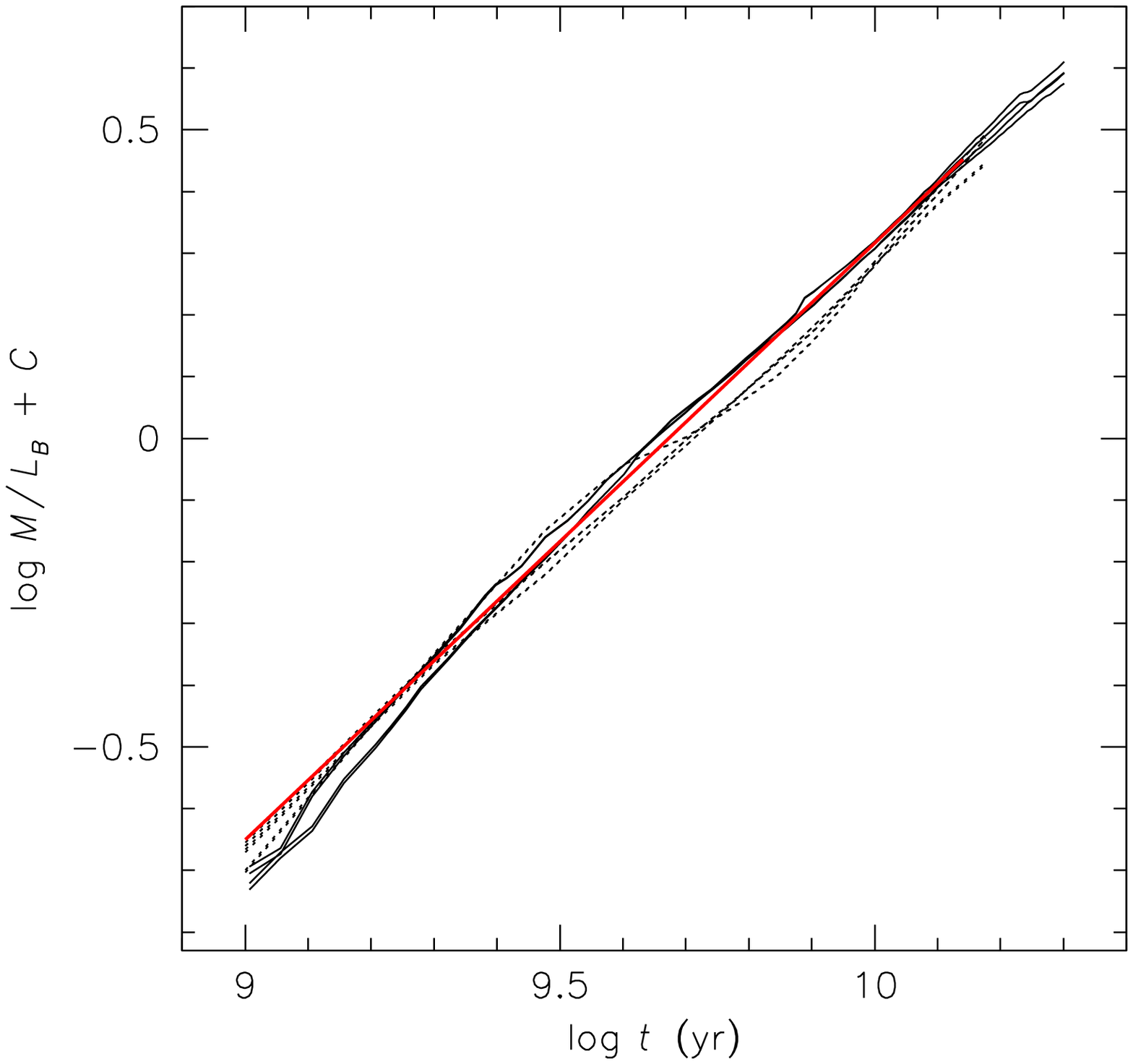}}
\figcaption{\small
Evolution of the $M/L_B$ ratio of a single age stellar population,
with arbitrary scaling. Solid lines are Bruzual \& Charlot (2003)
models with different metallicities and IMFs. Broken lines are
Maraston (2005) models with different metallicities, IMFs, and
implementation of horizontal branch stars. The predicted evolution
is very similar in all models. The red line is the
best-fitting powerlaw to a Solar metallicity, Salpeter (1955)
IMF Bruzual \& Charlot (2003) model.
\label{modfit.plot}}
\end{center}}

As first shown by {Tinsley} (1980)
the predicted evolution of the $M/L$ ratio of a single-age
stellar population
can be approximated by a powerlaw of the form
\begin{equation}
M/L \propto (t-t_*)^{\kappa},
\label{lumevo.eq}
\end{equation}
with $t_*$ the formation time of the stars. It can be deduced
from the data shown in
Fig.\ \ref{modfit.plot} that for any individual model
this approximation is accurate
to a few percent over the age range $9<\log t <10$.
Fitting powerlaws to all these models gives values of
$\kappa$ in the range 0.93--1.01.
A Solar metallicity
{Bruzual} \& {Charlot} (2003) model with a {Salpeter} (1955) IMF gives
a value of 0.97, and in the following we take $\kappa
=0.97 \pm 0.04$. We note that the synthetic $B$ filter adopted by
Bruzual \& Charlot (the Buser \& Kurucz [1978] $B_2$ filter)
is very similar to the Bessell (1990) $BX$ filter which
we use.

As emphasized by {Franx} (1993)
and {van Dokkum} \& {Franx} (2001) the observed evolution of
the $M/L_B$ ratio may underestimate the true evolution because of
selection effects. If galaxies undergo morphological evolution
and transform from late-type galaxies into early-type galaxies
at moderate redshift (see, e.g., {Dressler} {et~al.} 1997) the youngest
progenitors of today's early-type galaxies drop out of the sample
at high redshift. This ``progenitor bias'' leads to biased age
estimates, as we trace only the oldest galaxies and not the full
population of all progenitor galaxies. The significance of the
effect can be estimated from the observed evolution of the
early-type galaxy fraction in clusters and from the observed
scatter in the color-magnitude relation and FP (see {van Dokkum} \& {Franx} 2001).
The maximum effect on the evolution of the $M/L_B$ ratio occurs
if late-type galaxies are continuously transformed into early-type
galaxies {\em and} the scatter in the color-magnitude relation is entirely
due to age variations. In that extreme case the
observed luminosity evolution underestimates the true evolution by
$\Delta_{\rm pbias} \log (M/L_B) \sim 0.1z$. The
true effect is probably smaller than the maximum value, especially given recent
evidence that the evolution of the early-type
galaxy fraction out to $z\sim 1$ is weakest for the most massive objects
(Holden et al.\ 2006). We conservatively
assume $\Delta_{\rm pbias} \log (M/L_B) = (0.05 \pm 0.05) z$,
which encompasses the full range of possibilities.

With these assumptions and approximations
the only free parameter is the time of formation
of the stars $t_*$ (or the corresponding redshift $z_*$).
We determined the most likely value and its associated uncertainty
using Monte Carlo simulations. In each simulation,
each data point listed in Table 3
was perturbed by a value drawn from
a Gaussian distribution with a dispersion equal to the
uncertainty. Next, progenitor bias was taken into account by
decreasing all the measured $\log (M/L_B)$ ratios by $p \times z$,
with $p$
drawn from a top-hat probability distribution bounded by zero
and $0.1$. Finally, Eq.\ \ref{lumevo.eq}
was fit to the perturbed data to obtain $z_*$,
with the value of $\kappa$
drawn from a top-hat distribution bounded by $0.93$ and $1.01$.
From 1000 simulations we find that the mean luminosity weighted
formation redshift of the stars in massive cluster early-type
galaxies $z_* = 2.01^{+0.22}_{-0.17}$, where the uncertainties
indicate the 68\,\% confidence interval.
The best fitting model is shown in Fig.\ \ref{mlfit.plot}.
Ignoring progenitor bias (i.e., setting $p=0$)
gives $z_* = 2.23^{+0.24}_{-0.18}$.

\vbox{
\begin{center}
\leavevmode
\hbox{%
\epsfxsize=8.5cm
\epsffile{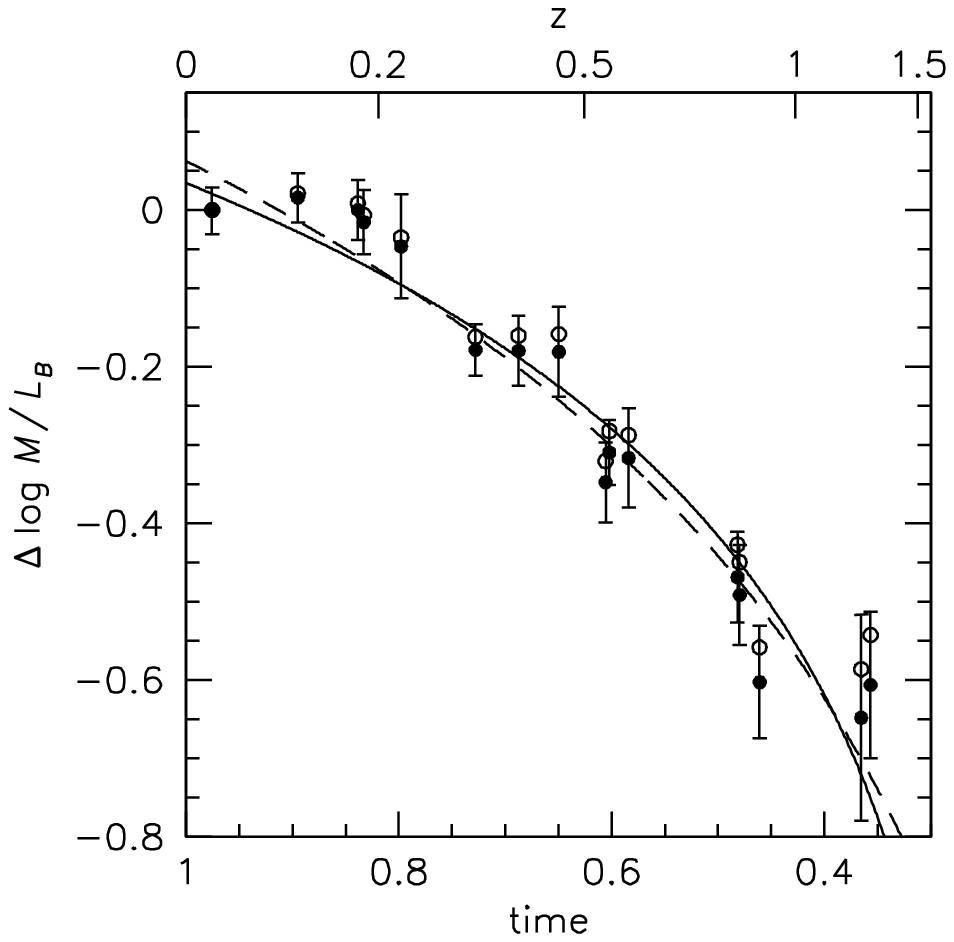}}
\figcaption{\small
Evolution of the mean $M/L_B$ ratio of massive cluster
galaxies with time. Open symbols are the same datapoints as
shown in Fig.\ \ref{mlz.plot}. Solid symbols with errorbars
are offset by $-0.05 \times z$ to account for progenitor bias
(see text). The solid line shows the best fitting model for a 
Salpeter-like IMF, which has a formation redshift of the stars $z_* = 2.01$. 
The broken line shows a model with a top-heavy IMF (slope
$x=0$) and a formation redshift $z_*=4.0$ (see \S\,\ref{conclusions.sec}).
\label{mlfit.plot}}
\end{center}}

\subsection{Comparison to Field Galaxies}
\label{field.sec}
\subsubsection{Data}

Several recent studies
of the fundamental plane of field early-type galaxies have found
that the most massive field galaxies out to $z\sim 1$
have similar $M/L$ ratios as the most massive cluster galaxies
(e.g., {Treu} {et~al.} 2005a; {van der Wel} {et~al.} 2005).
However,
{di Serego Alighieri}, {Lanzoni},  \& {J{\o}rgensen} (2006) find that massive cluster galaxies
are much older than massive field galaxies, based
on a comparison of data from {Treu} {et~al.} (2005a) (for field galaxies)
and {J{\o}rgensen} {et~al.} (2006) (for cluster galaxies) at $z\sim 0.9$.
Here we quantify the difference
between field and cluster early-types with 
$M>10^{11}$\,\msun\ using our sample of fourteen distant clusters
and self-consistent modeling of the field and cluster data.

The field samples of {van Dokkum} {et~al.} (2001), {van Dokkum} \& {Ellis}
(2003), Treu et al.\ (2005b),
and {van der Wel} {et~al.} (2005) are used.
Details are given in Appendix \ref{fielddata.sec}.
Masses and offsets in $M/L_B$ ratio
were calculated in the same way as for the cluster samples
(see \S\,\ref{procedure.sec}). The sample comprises 87 galaxies with
$M>10^{11}$\,\msun, ranging in redshift from 0.32
to 1.14. The sample was divided in bins of 10 galaxies
(with the lowest and highest redshift bins containing 8 and 9
galaxies respectively), and the
central redshift and $M/L_B$ ratio of each bin were determined using
the biweight estimator. The datapoints are listed in
Table 3 and shown in Fig.\
\ref{field.plot}, along with the cluster data (not corrected
for progenitor bias). The low redshift point is
determined from the {Bernardi} {et~al.} (2003a) sample
(see Appendix \ref{lit.sec}). The $M/L$ ratios of massive
field and cluster galaxies are very similar.

\subsubsection{Modeling Approach}

The $M/L$ evolution of field and cluster galaxies can
be described by
\begin{eqnarray}
(M/L)_{\rm clus} &=& A_c (t-t_c)^{\kappa_c}\\
(M/L)_{\rm field} &=& A_f (t-t_f)^{\kappa_f}
\end{eqnarray}
The standard approach for determining the star formation epoch
of field galaxies is the same as for cluster galaxies: the
rate of evolution of the $M/L$ ratio
is used to determine the luminosity-weighted star formation epoch $t_f$,
and $A_f$ is a
free parameter in the fit (see, e.g., {Treu} {et~al.} 2002, 2005a, 2005b; {van der Wel} {et~al.} 2005; {di Serego Alighieri} {et~al.} 2005). However, this approach
is not self-consistent when the evolution of field galaxies
is compared to that of cluster galaxies.
If field galaxies are younger than
cluster galaxies, their mean $M/L$ ratio evolves faster {\em and}
is offset from that of cluster galaxies. A self-consistent model
which describes the evolution of field and cluster galaxies
has $A_f \equiv A_c$, and has three rather than four
free parameters ($t_c$, $A_c$, and $t_f$).

The contrast between self-consistent models and standard models
is illustrated in Fig.\ \ref{fitdemo.plot}. The standard approach
(broken blue lines)
allows models which have different formation redshifts for field
and cluster galaxies to go through the same point at some
redshift, in practice $z\sim 0$ (see, e.g., {Treu} {et~al.} 2005b).
The implication is that
the difference in $M/L$ ratio
between field and cluster galaxies changes sign at $z\sim 0$. This
is not impossible, as it may be that field galaxies have a different
dust content, IMF, metallicity, or progenitor bias than cluster
galaxies. However, this would mean that we
live at a very special time, namely the only time in the past
or future of the universe when one or more of these effects
exactly cancel the difference in $M/L_B$ ratio resulting from the age
difference between field and cluster galaxies. It would also imply that
$\kappa_f \neq \kappa_c$, as differences in $A$ would in most models
be accompanied by  differences in the rate of evolution.
The approach followed in this paper is illustrated by the 
solid blue lines in Fig.\ \ref{fitdemo.plot}: differences in age result
in a different rate of evolution {\em and} an offset in $M/L$
ratio, such that the models converge at $t=\infty$ rather than
intersect at the present age of the Universe.

\vbox{
\begin{center}
\leavevmode
\hbox{%
\epsfxsize=8.5cm
\epsffile{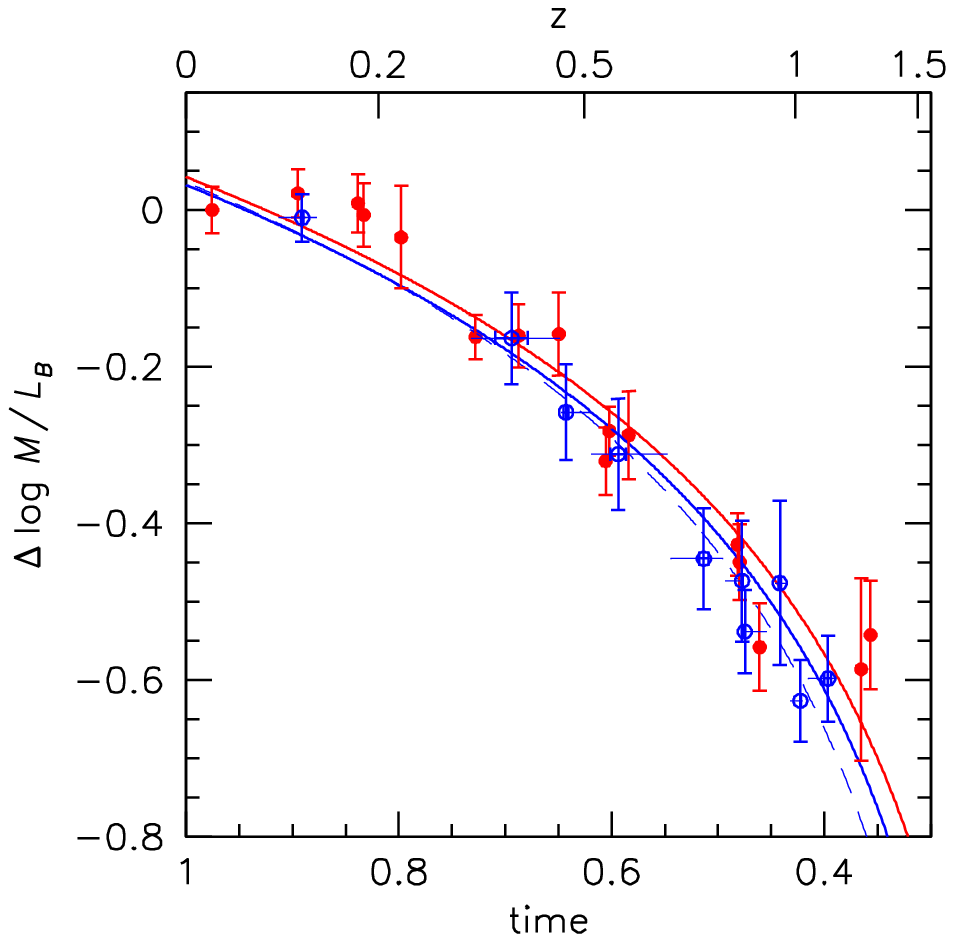}}
\figcaption{\small
Evolution of the mean $M/L_B$ ratio of field galaxies with
$M>10^{11}$\,\msun\ from the literature
(blue open symbols) compared to that of cluster galaxies
with $M>10^{11}$\,\msun\ (red solid symbols). Thin horizontal
lines show the range of redshifts covered in each bin.
No corrections
for progenitor bias were applied.
The red line shows the best fitting model to the cluster
galaxies. The broken blue line shows a fit to the field
galaxies only, and the solid
blue line shows a self-consistent
fit to the field galaxies given a model for the cluster galaxies
(see \S\,\ref{field.sec}). The age difference between massive field and
cluster galaxies is small at $\sim 4$\,\%. 
\label{field.plot}}
\end{center}}

\null
\vspace{-1cm}

\vbox{
\begin{center}
\leavevmode
\hbox{%
\epsfxsize=8.5cm
\epsffile{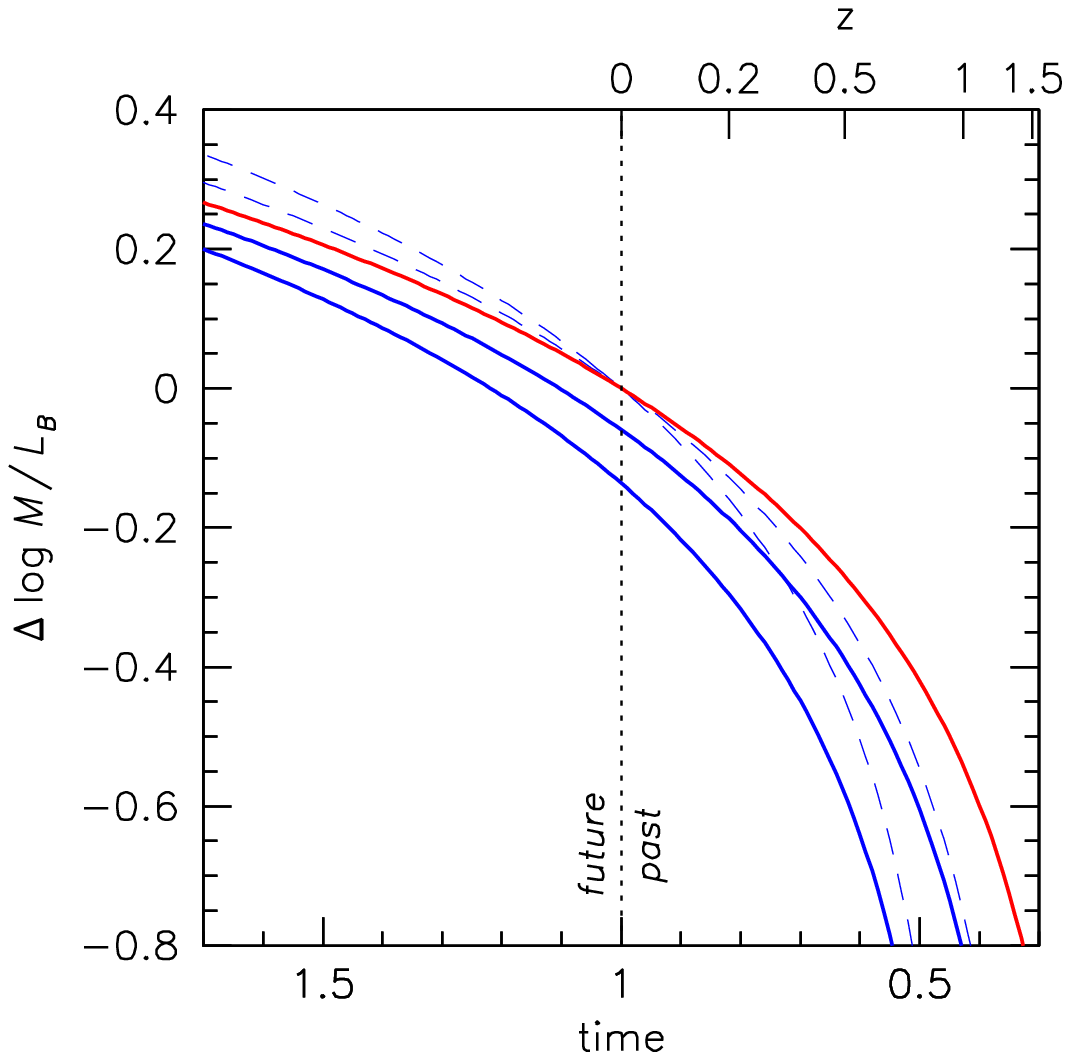}}
\figcaption{\small
Illustration of different ways to model the evolution of the
$M/L_B$ ratio of field galaxies. The red line shows the evolution
of a stellar population formed at $z_*=2.23$, appropriate for
cluster galaxies. Blue lines show the evolution of field
galaxies with $z_*=1.5$ (top)
and $z_*=1$ (bottom) respectively. Broken blue
lines illustrate the standard fitting approach, which allows
models to intersect at $z=0$.
Solid blue lines show the self-consistent
modeling approach adopted in the present study.
\label{fitdemo.plot}}
\end{center}}

\subsubsection{Results}

We determine the star formation epoch of field galaxies $t_f$
given a model for cluster galaxies ($t_c$ and $A_c \equiv A_f$).
For simplicity we
ignore progenitor bias:
the relative age of cluster and field galaxies is insensitive
to this bias as long as it is similar for
both populations. Taking $z_c=2.23$ for cluster galaxies
and fixing the zeropoint of the model
we find $z_f=1.95^{+0.10}_{-0.08}$ for field galaxies,
or an age difference of $4$\,\% at the present day. Note that
the quoted error does not include the uncertainty in the fit to
cluster galaxies.
Changing the amount of progenitor bias leads to negligible changes
in this result, as long as the bias is the same for field and
cluster galaxies.

The significance of this difference is assessed using Monte
Carlo simulations, which take the uncertainty in the
fit to the cluster galaxies into account. In each simulation
the $M/L_B$ ratios of cluster and field galaxies were perturbed
by values drawn from
Gaussian distributions with dispersions equal to the
uncertainties. Next the formation redshift of cluster galaxies
was determined, with the zeropoint of the model a free
parameter in the fit. The formation redshift of field
galaxies was also determined, with the zeropoint of the model
fixed to the value that was derived from the cluster galaxies.
Finally, the age difference was calculated from the formation
redshifts. From 1000 simulations we find that the age
difference between massive field and cluster galaxies is
4.1\,\%\,$\pm$\,2.0\,\%. Cluster galaxies are older
than field galaxies in 98\,\% of the simulations.

For completeness, we also determined the star formation epoch
of field galaxies using the standard approach, i.e., with
$A_f$ as a free parameter. As expected from the curves
in Fig.\ \ref{fitdemo.plot} the best-fit formation
redshift is lower: $z_f = 1.85^{+0.18}_{-0.13}$. This
model is shown by the broken blue line in Fig.\ \ref{field.plot}.

\section{Discussion and Conclusions}
\label{conclusions.sec}

In this paper we presented new measurements of kinematics and
structural parameters in three distant galaxy clusters.
In vdMvD06a,b the
spatially-resolved photometric and
kinematic profiles of the galaxies are
used to measure the evolution of the $M/L$ ratio independent
of many assumptions that enter the FP analysis. In the
present study we focused on the FP, and the implications
for the star formation epoch of the most massive galaxies
in clusters and in the field.

We find that the stars in massive early-type galaxies in clusters
have a mean
luminosity weighted formation redshift $z_*= 2.01^{+0.22}_{-0.17}$.
The implied ages are somewhat younger than most previous studies, which were
based on smaller samples (e.g., {Holden} {et~al.} 2005).
The quoted error reflects a combination of uncertainties
in the data points, the
stellar population synthesis models, and the significance of
progenitor bias. In particular, varying the amount of
progenitor bias from zero to the maximum
allowed by the {van Dokkum} \& {Franx} (2001) models [$\sim 0.1 z$ in
$\Delta \log (M/L_B$)] leads to a range in $z_*$ of $2.23-1.84$.
A special form of progenitor bias is ``dry'' merging, i.e.,
(nearly) dissipationless mergers of early-type galaxies
(e.g., {Tran} {et~al.} 2005b; {van Dokkum} 2005; {Bell} {et~al.} 2006; {De Lucia} {et~al.} 2006).
Numerical simulations suggest that
these mergers preserve the edge-on projection of
the FP relation
(e.g., {Gonz{\' a}lez-Garc{\'{\i}}a} \& {van Albada} 2003; {Boylan-Kolchin}, {Ma}, \&  {Quataert} 2006; {Robertson} {et~al.} 2006), which would imply that
our analysis is insensitive to this effect.

The main source of uncertainty that our analysis does not (fully)
take into account is the form of the IMF.
The rate of evolution, as parameterized by $\kappa$ in our
analysis, is very dependent on the logarithmic slope of the
IMF in the range $1-2$\,$M_{\odot}$.
{Tinsley} (1980) finds $\Delta \kappa \sim -0.3 \Delta x$,
where $x$ is the slope of the IMF, and the more recent
{Worthey} (1994) models give $\Delta \kappa \sim -0.22 \Delta x$.
Although we explored the values of $\kappa$ for most commonly used
IMFs ({Salpeter} 1955; {Kroupa} 2001; {Chabrier} 2003)
these are all very similar 
in the region around one Solar mass. These IMFs are probably
appropriate for star formation in present-day
disk galaxies, but may not be
applicable to (the progenitors of) early-type galaxies.
As is well known the large mass of heavy elements observed in the
hot gas of massive clusters, the abundance ratios of
early-type galaxies, numerical simulations of star formation at very
high redshift, and other lines of evidence suggest that
the IMF may have been top-heavy at early times
(see, e.g., {Worthey}, {Faber}, \&  {Gonzalez} 1992; {Larson} 1998; {Schneider} {et~al.} 2002, and references therein).
The top-heavy IMF preferred by {Nagashima} {et~al.} (2005) has $x=0$ (where
$x=1.35$ is the {Salpeter} [1955] slope), which implies
$\kappa_B = 1.4$. This value of
$\kappa$ gives a best fitting formation redshift of
$z_* = 4.0$, higher than the $z \sim 2$ range derived for
Salpeter-like IMFs. This model is shown by the broken line in Fig.\
\ref{mlfit.plot}. We note that
our observations probe a very different range of stellar
masses than the abundance studies
($1-2$\,\msun\ versus $\sim 10$\,\msun), and that
a top-heavy IMF does not necessarily
imply an overall change in the slope but could instead
be caused by a time-varying characteristic mass
({Larson} 1998).

Interestingly, there is some indirect
evidence for a top-heavy IMF from the combination of our data with
recent results for galaxies at higher redshifts.
Although many massive galaxies at $z\sim 2.5$
are still in the process of forming stars (see \S\,1),
there is also a population
with very low specific star formation rates and
strong Balmer/4000\,\AA\ breaks (e.g.,
Kriek et al.\ 2006a,b). These objects formed most of their stars at
redshifts $z\sim 3$ or beyond, inconsistent with the star
formation epoch that we derive
for Salpeter-like IMFs but in agreement with our results
for a top-heavy IMF. Alternative explanations are also possible;
e.g., the apparently old $z\sim 2.5$ galaxies may be
``rejuvenated'' at later times (see, e.g., Treu et al.\ 2005b),
or they could simply be the oldest objects in a
wide distribution of progenitor galaxies.

The age difference between massive field and cluster galaxies is better
constrained than their absolute ages, as it 
is less affected by the uncertainty in the IMF.
Our self-consistent modeling approach gives different constraints than
the standard approach, as it has less free parameters. Assuming that
progenitor bias is similar for massive cluster and field galaxies,
they have a Salpeter-like IMF, and they have the same
metallicity and dust content at a given mass, we find
that massive field galaxies are younger
than massive cluster galaxies, but that the difference is small
at $4.1 \pm 2.0$\,\% (or $0.4 \pm 0.2$ Gyr).
If a top-heavy IMF with $x=0$ is assumed for both field and
cluster galaxies the age difference is 3.0\,\%\,$\pm$\,1.5\,\%.

We note that the quoted errors do not include systematic errors
in the comparison of field and cluster galaxies, which are
difficult to quantify. Possible sources
of such errors are differences in dynamical structure that deviate
from pure homology,
selection effects induced by the
magnitude limits of the field samples (although these effects
are more relevant for low mass galaxies; see Treu
et al.\ 2005, van der Wel et al.\ 2005), a
difference in $\kappa$ and/or $A$ (which would reflect differences in the
IMF or the metallicity), and a difference in the amount
of progenitor bias. We tested the effect of a difference in
progenitor bias by assuming that cluster early-type galaxies are not
affected by it whereas the $M/L$ ratios of field early-type galaxies need to
be corrected by $-0.1 \times z$. In that case the age difference
increases by a factor of two, to $\approx 9$\,\%. The importance
of progenitor bias is bounded by the
observed scatter in
the color-magnitude relation of early-type galaxies
(van Dokkum \& Franx 2001). In clusters
this scatter remains low all the way to $z\sim 1.3$
(Mei et al.\ 2006), but it is not yet clear whether this
also applies to massive field early-type galaxies
(see, e.g., Ferreras et al.\ 2005).

The small age difference we find is consistent with most, but not
all, previous studies. It is consistent with the high
redshift field FP studies of {Treu} {et~al.} (2005b) and {van der Wel} {et~al.} (2005), the analysis of a sample of gravitational lenses by
Rusin \& Kochanek (2005),
and with studies of the local Mg$_2$ -- $\sigma$ relation
and FP by {Bernardi} {et~al.} (1998, 2003a, 2006).
However, other studies find larger age differences (always in the
sense that field galaxies are younger than cluster galaxies).
For nearby massive early-type
galaxies, {Thomas} {et~al.} (2005) and {Clemens} {et~al.} (2006) find age 
differences of 1.5 -- 2 Gyr from complex models which include age,
metallicity, $\alpha$-enhancement, and (in the case of
Clemens et al.) Carbon-enhancement as free parameters.
Such large age differences are inconsistent with our analysis
at $>3\sigma$.
Perhaps most importantly, our results are
not in agreement with the recent study of {di Serego Alighieri} {et~al.} (2006),
who compare the $M/L$ ratios of field galaxies at $z\sim 1$ from
{Treu} {et~al.} (2005b) and {di Serego Alighieri} {et~al.} (2005) to that of galaxies in two clusters
with mean redshift $\langle z\rangle
= 0.86$ from {J{\o}rgensen} {et~al.} (2006).
For galaxies with masses $>10^{11}$\,\msun\
{di Serego Alighieri} {et~al.} (2006) find a very large age difference of 3.5 -- 4 Gyr
(their Fig.\ 4). The source of the
discrepancy between their results and ours is, at least in part,
due to the cluster data that are used in the comparison to field galaxies.
As discussed in Appendix \ref{lit.sec} the
$M/L$ ratios of the galaxies
in the {J{\o}rgensen} {et~al.} (2006) study need to be corrected
downward by a factor $1+z$. Applying this offset to the
$\langle z \rangle = 0.86$ cluster
galaxies in Figs.\ 1 and 2 of di Serego Alighieri et al.\ (2006) brings
them in line with the $z\sim 1$ field galaxies.\footnote{See
also the erratum published by di Serego Alighieri et al.\ (2006)
in response to our discovery of this discrepancy. Their revised
conclusions are consistent with those presented here.}

The small age difference between massive field and cluster
early-type galaxies that we find is remarkable
in the context of ``standard'' hierarchical models, which had predicted
that early-type galaxies in clusters have much older
stellar populations than those in the general field
(e.g., {Baugh}, {Cole}, \& {Frenk} 1996; {Kauffmann} 1996; {Kauffmann} \& {Charlot} 1998).
Specifically, as shown in van Dokkum et al.\ (2001), the models of
Kauffmann et al.\ (1999) and Diaferio et al.\ (2001) predicted a
systematic offset between field and cluster
galaxies of $\sim 0.25$ in $\log (M/L_B)$ at all redshifts, clearly
at odds with observations.
The reason for this
environmental dependence in these models is that gas cooling and
subsequent star formation continues as long as a galaxy is
the central galaxy in its halo. Star formation terminates
when a galaxy becomes a satellite in a larger halo,
which naturally leads to a large population
of ``red and dead'' galaxies in clusters and blue star forming
galaxies in the field.
This inability of standard models to produce red field galaxies
is well known, and among proposed solutions is to
prevent cooling by postulating that the gas is heated by a central active
galactic nucleus (AGN) (e.g., {Bower} {et~al.} 2006; {Kang}, {Jing}, \& {Silk} 2006; Croton et al.\ 2006).
The recent study of {De Lucia} {et~al.} (2006) incorporates such AGN feedback
in a very large cosmological simulation.
In their simulation, the
median star formation epoch $z_* \sim 2.6$ for elliptical galaxies with masses
$M>10^{11}$\,\msun, in good agreement with our results. Furthermore,
judging from their Fig.\ 7 the difference in age
between ellipticals in halos of $10^{13}$\,\msun\ and $10^{15}$\,\msun\
is only $\sim 0.7$ Gyr, consistent with our study.
We note that this success does not necessarily imply that
the models can reproduce the observed evolution of the FP.
Building on the semi-analytical models of Baugh et al.\ (2005) and
Bower et al.\ (2006), Almeida,
Baugh, \& Lacey (2006) find that evolution in the radii of
early-type galaxies compensates for the evolution in their $M/L$ ratios,
such that the zeropoint of the FP is approximately constant
with redshift -- inconsistent with the observations.

Taken at face value, the age difference between field and cluster
galaxies is $\sim 0.4 \pm 0.2$\,Gyr, which should be fairly easily
detectable at $z\gtrsim 2$. Interestingly, this number
agrees reasonably
well with the age difference that {Steidel} {et~al.} (2005)
find comparing Lyman break
galaxies in the general field to those in a proto-cluster at $z=2.3$.
It is also qualitatively consistent with the finding that red
galaxies at $z\sim 2.5$ cluster more strongly than blue galaxies
(Quadri et al.\ 2006). 

\begin{acknowledgements}

We thank Arjen van der Wel and Brad Holden for their sleuthing,
and the referee, Inger J\o{}rgensen, for her comments and generous
sharing of unpublished data.
Support from NASA grants HF 01126.01-99A,
HST AR-09541.01-A, and LTSA NNG04GE12G, and
from National Science Foundation
grant NSF CAREER AST 04-49678 is gratefully acknowledged.
The authors wish to recognize and acknowledge
the very significant cultural role and reverence that the summit of
Mauna Kea has always had within the indigenous Hawai'ian community. We
are most fortunate to have the opportunity to conduct observations
from this mountain.

\end{acknowledgements}



\begin{appendix}

\section{Literature Data for Cluster Galaxies}
\label{lit.sec}

In addition to the three clusters discussed in this paper we used eleven
additional distant clusters in the analysis. We also included
the Coma cluster and a sample of SDSS galaxies in the nearby
universe. Here we describe the various transformations that were
applied to bring the published data to our system. We also describe
sources of systematic uncertainty for each cluster. In most cases
morphological information is available; only early-type galaxies
with masses $>10^{11}$\,$M_{\odot}$ were included in the calculation
of the $M/L$ offsets.

\subsection{Coma (Abell 1656)}
\label{coma.sec}

The value of $c_0$ in Eq.\ \ref{dml.eq} is set to that of the
nearby Coma cluster. This choice is arbitrary, as adding a constant
to all values of $\Delta \log (M/L_B)$ does not change our results
in any way. We use the sample of {J{\o}rgensen} {et~al.} (1996). Structural parameters
measured in the $B$ band are given in J{\o}rgensen {et~al.} (1995a), and
velocity dispersions corrected to a $1\farcs 7$
radius aperture are listed in {J{\o}rgensen} {et~al.} (1995b).
The listed surface brightnesses are corrected from the
average brightness within the effective radius $\langle \mu
\rangle_e$ to that
at the effective radius: $\mu_e = \langle \mu \rangle_e + 1.393$.
Effective radii are converted from arcseconds to kpc by adopting
a Hubble flow velocity $v_{\rm flow} = 7376 \pm 223$\,\kms\
(see vdMvD06b).

Restricting the sample to early-type galaxies with $M>10^{11}$\,\msun\
we find $c_0 = -9.626$. The value of $\Delta \log (M/L_B)$ is zero
by definition, with a random uncertainty of $\pm 0.014$.
The systematic uncertainty in $\log (M/L_B)$ is a combination
of several factors. Zeropoint uncertainties in the $B$ band photometry
give $\pm 0.010$ ({J{\o}rgensen}, {Franx}, \&  {Kj\ae{}rgaard}
1992), and the uncertainty in
$v_{\rm flow}$ implies $\pm 0.013$. A comparison of studies
by different authors gives $\pm 0.02$ (see {Hudson} {et~al.} 2001, and
vdMvD06b);
this uncertainty includes (and may be dominated by)
systematic differences in the methodology for deriving
structural parameters and velocity dispersions.
As discussed in vdMvD06b
several other systematic uncertainties cancel in the comparison to
the distant clusters. For example,
the uncertainty in the Hubble constant cancels
because we are only concerned with the evolution of the $M/L_B$
ratio and not with the absolute value. Also, as discussed in
{van Dokkum} \& {Franx} (1996) our methodology for measuring velocity dispersions
of galaxies in distant clusters
mimics that of J{\o}rgensen et al.\ in the nearby
universe. The combined systematic error is $0.026$, and
assuming that the random and systematic errors can be added
in quadrature we obtain $\Delta \log (M/L_B) = 0.000 \pm 0.029$.

\subsection{The Bernardi Sample}

{Bernardi} {et~al.} (2003a) have studied the fundamental plane
relation of galaxies in the SDSS. This sample has
several advantages over previous studies: it is large
(the total sample comprises 8661 objects), does
not suffer from uncertainties due to peculiar velocities
(both because they average out over the large volume
probed by the data and because of the
relatively high typical redshift of $0.1$), has very
homogeneous spectroscopy and multi-band photometry, probes
a large range of environments, and spans a
range in redshift so that in principle
evolutionary effects can be studied within the sample.
An important drawback specific to this sample is that the
magnitude-selection causes significant biases which
need to be taken into account ({Bernardi} {et~al.} 2003a); such
biases are less important (although not absent) for cluster
samples as all galaxies are at the same distance.
Furthermore, as noted by, e.g., {Cappellari} {et~al.} (2006) the
coefficients $a$ and $b$ that Barnardi et al.\ derive do not
appear to be consistent with most other studies of the FP;
the cause for this discrepancy is unclear at present.

Velocity dispersions and structural parameters in the
rest-frame $g'$ band were obtained from Table 3b
of {Bernardi} {et~al.} (2003b). Effective radii were
converted to our cosmology. Rest-frame $g'$ band
surface brightnesses were converted to the rest-frame $B$ band using
the transformation
\begin{equation}
\label{trafob.eq}
B = g' + 0.44 (g'-r') + 0.17,
\end{equation}
with $B$ on the Vega system and $g'$ and $r'$ on the AB system.
For consistency this transformation was derived using the same method
and $B$ band filter curve
as Eq.\ \ref{trafo.eq}, with the only difference that the redshift
was set to zero (i.e., the $g'$ and $r'$
filters were not redshifted). {Fukugita} {et~al.} (1996) give
$B_z = g' + 0.42 (g'-r') + 0.22$; for the typical colors
of galaxies in the sample this transformation
is consistent with ours to $0.04$ mag. The $g'-r'$ colors
cannot be obtained by simply taking the difference $\langle \mu_{g'}
\rangle_e
-\langle \mu_{r'}\rangle_e$ as the surface brightness was evaluated
within a different radius in each band. Therefore we first corrected the
surface brightnesses to a common radius using the
measured effective radii of the galaxies (see {van der Wel} {et~al.} 2005).
Finally we applied the transformation $\mu_e = \langle \mu
\rangle_e + 1.393$.
We note that the uncertainty in this procedure is ultimately
determined
by the difference between the observed band and the final rest-frame
band. In this case, the observed band is the $g'$ band and the
final rest-frame band is the $B$ band, and at $z\sim 0.1$ these are very
well matched.

{Bernardi} {et~al.} (2003b) corrected their velocity dispersions to
an aperture of radius $0.125 r_e$, which for almost all
galaxies is significantly smaller than the $1\farcs 7$
aperture at the distance of Coma that we adopted.
We first retrieved the dispersions as measured
through the $1\farcs 5$ SDSS fibers using the measured
effective radii (in arcsec) and Eq.\ 1
of {Bernardi} {et~al.} (2003b). Then we corrected these measured
dispersions to a $1\farcs 7$ aperture at the distance of Coma
using the redshifts of the galaxies and the prescription of
{J{\o}rgensen} {et~al.} (1995b).

FP offsets were determined for all galaxies in the sample
with local density $n_{\rm den}>20$, using
$a=1.20$, $b=-0.83$, and $c_0=-9.626$. Following
{Bernardi} {et~al.} (2003b) (their Fig.\ 10)
and {Bernardi} {et~al.} (2003a) (their Fig.\ 9)
galaxies with $n_{\rm den}=100$ were excluded.
As discussed extensively
in {Bernardi} {et~al.} (2003a) average $M/L$ ratios inferred from
these data need to be corrected for selection effects.
Fitting to the values listed in their Fig.\ 7 gives
$\Delta_{\rm select} \log (M/L_{g'}) = -0.289 z$ and $\Delta_{\rm select}
\log (M/L_{r'}) = -0.208 z$. Extrapolating to the $B$ band using
Eq.\ \ref{trafob.eq} gives $\Delta_{\rm select} \log (M/L_B) = -0.320 z$.
Bernardi et al.\ determined the effect for the full sample, and
not for the subsample of galaxies with $M>10^{11}$\,\msun\ that
is relevant here. Selection effects will be less important for the
massive galaxies, but it is not clear by
how much. We assume that the effect is half that of
the full sample, and conservatively assign an uncertainty which
encompasses the full range of possibilities:
\begin{equation}
\Delta \log (M/L_{B,\rm corr}) = \Delta \log (M/L_{B}) + (0.16 \pm 0.16) z
\end{equation}
Given the uncertainty in the redshift evolution within the Bernardi
et al.\ sample we evaluate the average $M/L_B$ offset at one redshift
only. We chose $z\approx 0.10$, as selection effects do not play a large role
at that redshift and the
local galaxy densities $n_{\rm den}$ are not well determined for
$z\lesssim 0.08$ ({Bernardi} {et~al.} 2003b). In practice we select
the 171 galaxies with $M>10^{11}$\,\msun, $20<n_{\rm den}<100$,
and $0.075<z<0.14$. The average redshift of this sample is $z=0.109$.

The value of the $M/L$ offset is listed in Table 3.
The random uncertainty is only $\pm 0.008$. The uncertainty
in the photometric transformation is taken to be the difference
between the {Fukugita} {et~al.} (1996) transformation and ours ($\pm 0.016$).
The uncertainty due to selection effects is $0.16 \times 0.109 = 0.017$.
{Bernardi} {et~al.} (2003b) estimate that the systematic uncertainty in the
velocity dispersions is $\sim 3$\,\%, implying $\pm 0.019$ in $\log
M/L_B$. Adding all uncertainties in quadrature gives $\pm 0.030$
for the total error in the $M/L$ offset.

For the analysis in \S\,\ref{field.sec} we also determined the
$M/L$ offset in low-density regions, selecting all 1519 galaxies
with $M>10^{11}$\,\msun, $0.075<z<0.14$, and $n_{\rm den}<10$.
The average redshift of this sample is 0.1093. The offset (corrected
for selection effects)
is listed in Table 3. The random error is
$0.003$ and the systematic error is the same as derived
for the high-density sample. The difference between the high- and
low-density offset is $0.031$ in $\log (M/L_B)$, or 0.08 magnitudes
if it is interpreted as a difference in surface brightness at
fixed $r_e$ and $\sigma$. This difference appears to be
consistent with the trends shown in Fig.\ 9 of {Bernardi} {et~al.} (2003a)
for this redshift range.

\subsection{Abell 2218}

Abell 2218 is a well-studied very rich
cluster at $z=0.1756$ ({Le Borgne}, {Pello}, \&  {Sanahuja} 1992).
Its FP has been analyzed in two independent studies, {J{\o}rgensen} {et~al.} (1999)
and {Ziegler} {et~al.} (2001). {J{\o}rgensen} {et~al.} (1999)
give velocity dispersions and two sets of structural parameters, one
derived from HST images and one derived from ground-based data.
Their velocity dispersions have already been corrected to the
$1\farcs 7$ aperture at the distance of Coma. The ground-based
structural parameters were corrected from the observed $I_c$ band
to the rest-frame $B$ band using the listed $V-I$ colors:
\begin{equation}
\label{a2218.eq}
B_z = I_c + 1.29 (V-I_c) + 0.18,
\end{equation}
and corrected from $\langle \mu \rangle_e$ to $\mu_e$.
The structural parameters measured from the HST F702W images were
transformed by J{\o}rgensen et al.\ to the $I_c$ band, and we use
Eq.\ \ref{a2218.eq} to transform the surface brightness to rest-frame
$B$.

{Ziegler} {et~al.} (2001) obtained independent spectroscopy for galaxies in
Abell 2218, and give velocity dispersions which were corrected
to the same $1\farcs 7$ radius aperture at the distance of Coma
as we use. They used the same HST imaging as {J{\o}rgensen} {et~al.} (1999), but
derive their own structural parameters from the images.
We transformed their listed rest-frame Gunn $r$
surface brightnesses from $L_{\odot}$\,pc$^{-2}$
to mag\,arcsec$^{-2}$ using their Eq.\ 7, and then converted them
back to observed $I_c$ magnitudes using their Eq.\ 1. In this
conversion we used the listed $V$ and $I_c$ magnitudes after applying
extinction corrections of 0.083 mag in $V$ and 0.048 mag in
$I_c$. The $I_c$ surface brightnesses were transformed from
$\langle \mu \rangle_e$ to $\mu_e$ and corrected to
rest-frame $B$ using Eq.\ \ref{a2218.eq}. 

Combining all information there are three measurements of the
FP: 1) dispersions and HST structural parameters from {J{\o}rgensen} {et~al.} (1999);
2) dispersions and ground-based structural parameters from
{J{\o}rgensen} {et~al.} (1999); and 3) dispersions and HST structural parameters from
{Ziegler} {et~al.} (2001). The two sets of dispersions and the two sets
of ground-based data are independent. We determined the $M/L$ offset
for all three cases and compared the results. The offsets agree
very well: differences are approximately $0.02$ in $\log (M/L_B)$.
Rather than take the average of the three determinations we use
the zeropoint from measurement 2): it falls between
the other two and it is the one used in the analysis of the
{J{\o}rgensen} {et~al.} (1999) paper. Apart from a systematic uncertainty of 0.02
due to the differences between the three determination there is
also an uncertainty introduced by the fact that
all structural parameters were measured in redder bands than the
redshifted $B$ band. This uncertainty stems from color
gradients and the dependence of the FP parameters on passband,
and is estimated at 0.05 mag. The combined systematic
uncertainty is 0.029 in $\log (M/L_B)$.

\subsection{Abell 665}

The FP in Abell 665 ($z=0.1829$; {G{\'o}mez}, {Hughes}, \&  {Birkinshaw} 2000) was studied by
{J{\o}rgensen} {et~al.} (1999). As for Abell 2218 there are two sets of structural
parameters, one from HST imaging and the other from ground-based
imaging. Contrary to Abell 2218 there are also two sets of $V-I_c$
color measurements: one directly measured from the ground and
the other transformed from $R_{606}-I_{814}$ HST colors. We
applied the same transformations as for Abell 2218, and compared
the FP relations derived from ground-based data to the FP
derived from HST data.
The difference is $0.02$ in $\log (M/L_B)$,
consistent with the expected uncertainties due to the color
transformations. The final offset was calculated from the
HST measurements in J\o{}rgensen et al.\ (1999).
As for Abell 2218
the total systematic uncertainty is estimated at
$0.029$ in $\log (M/L_B)$.

\subsection{Abell 2390}

{Fritz} {et~al.} (2005) present the FP relation in Abell 2390 at $z=0.228$.
We corrected the listed velocity dispersions to our fiducial aperture
following {J{\o}rgensen} {et~al.} (1995b), using $1\farcs 5 \times 2\farcs 8$
as the extraction aperture. The listed rest-frame Gunn $r$
surface brightnesses were converted back to observed WFPC2
$I_{814}$ magnitudes using $\langle \mu_{814}\rangle_e =
\langle \mu_{r,z} \rangle_e - 0.75$
(see Fritz et al.\ 2005). We converted
$\langle \mu \rangle_e$ to $\mu_e$ and transformed to the
(COSMIC camera) $I$ band using
\begin{equation}
I = I_{814} - 0.037 (B-I) + 0.007 (B-I)^2 + 0.00
\end{equation}
({Holtzman} {et~al.} 1995). The $B-I$ colors were determined from
the listed $B$ and $I$ aperture magnitudes. These are
not corrected for extinction; we applied corrections of 0.476 mag
to $B$ and $0.214$ mag to $I$ (see {Fritz} {et~al.} 2005).
Rest-frame $B$ surface brightnesses were determined using
\begin{equation}
B_z = I + 0.47 (B-I) + 0.51.
\end{equation}
Based on listed zeropoint uncertainties the
uncertainty in the photometric transformations is estimated at
$0.03$ mag.

In contrast to our own measurements of the FP in distant clusters
no effort was made to determine parameters in the same way as
the J{\o}rgensen et al.\ studies at low redshift. There is therefore
a systematic uncertainty of $\approx 5$\,\% in the velocity
dispersions, which is caused by possible differences in the
fitting region, choice of templates, and fitting methodology.
This corresponds to an uncertainty of $0.031$ in $\log (M/L_B)$.
Combined with a 0.05 mag uncertainty due to the transformations
and the fact that the
FP was determined in red rest-frame bands, we estimate that
the total systematic uncertainty is 0.037 in $\log (M/L_B)$.

\subsection{CL\,1358+62}

{Kelson} {et~al.} (2000a, 2000b, 2000c) provide
an extensive analysis of the FP in the cluster CL\,1358+62 at
$z=0.327$ in the rest-frame $V$ band. Listed velocity
dispersions ({Kelson} {et~al.} 2000b) are already corrected to our
fiducial aperture. Listed
rest-frame surface brightnesses and colors were back-corrected
to the observed ones using the Eqs.\ in {Kelson} {et~al.} (2000a).
They were then transformed to rest-frame $B$ using
\begin{equation}
B_z = I_{814} + 1.09 (R_{606} - I_{814}) + 0.52
\end{equation}
and transformed from $\langle \mu \rangle_e$ to $\mu_e$.

The Kelson et al.\ methodology is essentially identical to ours,
and systematic uncertainties in, e.g., spectral continuum filtering
cancel. Systematic uncertainties in the dispersions
due to possible changes of the
spectral templates with redshift are estimated at 2\,\%.
Systematic errors in the photometric transformations are
estimated at 0.03 mag. The combined systematic uncertainty
is $0.021$ in $\log (M/L_B)$.

\subsection{CL\,0024+16}

{van Dokkum} \& {Franx} (1996) present the FP in the $z=0.391$ cluster
CL\,0024+16. We use their sample rather than the much larger sample
of {Moran} {et~al.} (2005) because {van Dokkum} \& {Franx} (1996) use identical techniques
to ours, and systematic uncertainties in the velocity dispersions
dominate over random uncertainties due to the limited sample size.
We note that {Moran} {et~al.} (2005) find that their data agree
with those of {van Dokkum} \& {Franx} (1996) to within a few percent, both for
the velocity dispersions of individual objects in common between
the two samples and for the offset in $M/L$ ratio derived from the FP.

Listed velocity dispersions have already been corrected to our
fiducial aperture. The listed $I$-band
surface brightnesses were corrected to
rest-frame $B$ using
\begin{equation}
B_z = I + 1.46 (R-I) + 0.57.
\end{equation}
The systematic uncertainty stems from the same sources as
for CL\,1358+62, and is estimated at 0.021 in $\log (M/L_B)$.

\subsection{MS\,2053--04 and MS\,1054--03}

{Wuyts} {et~al.} (2004) analyze the FP in the two clusters
MS\,2053--04 ($z=0.583$) and MS\,1054--03 ($z=0.832$). The Wuyts et al.\
study followed initial studies of smaller samples in the two clusters by
{Kelson} {et~al.} (1997) and {van Dokkum} {et~al.} (1998a) respectively. The listed
structural parameters and velocity dispersions are already on our
system. However, we update
the transformations to rest-frame $B$, as the synthetic 
$B$ filter curve
used in Wuyts et al.\ (2004) is slightly different from the one
used in the present study. The
updated transformations are
\begin{equation}
B_z = I_{814} + 0.38 (R_{606}-I_{814}) + 0.93
\end{equation}
for MS\,2053--04 and
\begin{equation}
B_z = I_{814} - 0.05 (R_{606}-I_{814}) + 1.22
\end{equation}
for MS\,1054--03. These transformations are consistent with those
given in Wuyts et al.\ (2004) to within $\approx 0.04$ mag for the
typical colors of early-type galaxies in these clusters.
The systematic uncertainty is taken
to be the same as for CL\,1358+62.

\subsection{RX\,J0152.7--1357}
\label{rxj0152}

Imaging and spectroscopic data of RX\,J0152--13 ($z=0.837$) and
RX\,J1226+33 ($z=0.892$) were obtained in the context
of the Gemini/{\em HST} Galaxy Cluster Project
(J\o{}rgensen et al.\ 2005).
J\o{}rgensen et al.\ (2006) discuss the FP in these clusters,
finding that the most massive galaxies in
these clusters are offset from the Coma FP by $\Delta
\log M/L_B \approx -0.25$ only. This offset is much smaller than
had been found previously for the cluster MS\,1054--03 at
$z=0.83$ (Wuyts et al.\ 2004), indicating significant
cluster-to-cluster scatter at $z\sim 1$.
J\o{}rgensen et al.\
(2006) do not provide data for individual galaxies in these
clusters, but for RX\,J0152--13 the FP can be constructed
by combining velocity dispersions listed in Table 12
of J\o{}rgensen et al.\
(2005) with structural parameters measured by Blakeslee
et al.\ (2006). The velocity dispersions ($\sigma_{\rm corr}$
from J\o{}rgensen et al.\ [2005]) are already on our
system, and no further corrections are required. Effective
radii from Blakeslee et al.\ were circularized by multiplying
them by $\sqrt{b/a}$, and converted to kpc. 
The listed
magnitudes ($i_{775, \rm gfit}$ in Blakeslee et al.\ [2006])
were converted to surface brightnesses on our system through
the transformation
\begin{equation}
\mu_{e, 775} = i_{775, \rm gfit} + 2.5 \log 2 + 2.5 \log
(\pi r_e^2) + 1.393,
\end{equation}
with $r_e$ in arcsec. Next, surface brightnesses
and colors were converted from AB to Vega magnitudes using the
Sirianni et al.\ (2005) ACS zeropoints, and transformed to
rest-frame $B$ using
\begin{equation}
B_z = i_{775} - 0.35 (i_{775} - z_{850}) + 1.21.
\end{equation}
Finally, the $M/L$ offset was derived using the same techniques
as for the other clusters. We find $\Delta \log M/L_B = -0.449$,
much larger than the J\o{}rgensen et al.\ (2006) offset.
This difference is caused by a combination of two effects:
small differences in methodology (e.g., a 0.05 mag difference
between the measured Blakeslee et al.\ [2005] and
J\o{}rgensen et al.\ [2006] $i_{775}$ surface brightnesses),
and an error in the transformation from observed to rest-frame
magnitudes. The rest-frame
magnitudes in J\o{}rgensen et al.\ (2006)
are too faint by a factor of $1+z$.\footnote{See the erratum
published by J\o{}rgensen et al.\ (2006) in response to our
discovery of this discrepancy.}
I.\ J\o{}rgensen kindly determined the offset for galaxies
with $M>10^{11}$\,\msun\ in RX\,J0152--13
using the J\o{}rgensen et al.\ (2006) data
with our transformation from SDSS $i'$ and $z'$
magnitudes to rest-frame $B$, and our
zeropoint for the Coma cluster. The offset is $-0.475$. This
value is consistent with ours to $<0.01$, when the systematic
difference of
0.05 mag between the Blakeslee et al.\
and J\o{}rgensen et al.\ methodology is taken into account.
Sources of systematic error are the same as for, e.g.,
Abell 2218 and Abell 665, but we added 0.02 in quadrature
to reflect the difference between the Blakeslee et al.\
and J\o{}rgensen et al.\ structural parameters.

\subsection{RX\,J1226.9+3332}

This $z=0.892$ cluster is discussed in J\o{}rgensen
et al.\ (2006) together with RX\,J0152--13.
No individual measurements of galaxies are
available, but I.\ J\o{}rgensen determined the offset in
$M/L_B$ ratio using our transformations, Coma zeropoint,
and mass selection. A small correction of 0.05 mag
was applied to bring the data to the same
system as the Blakeslee et al.\ (2006) photometry
(see \S\,\ref{rxj0152}).
The offset is consistent
with the value that is implied by
the J\o{}rgensen et al.\ (2006) study, after correcting
their data points by a factor $1+z$ (\S\,\ref{rxj0152}).
Sources of systematic uncertainty are the correction to the
Blakeslee et al.\ (2006) system (0.02 in $\log M/L_B$),
systematic uncertainty in the dispersions (0.025), the
transformation to rest-frame $B$ (0.012), and the
photometric zeropoints of the ground-based photometry
(0.02). The total systematic uncertainty is 0.040.
The random uncertainty is 0.039 in $\log M/L_B$, which
brings the total error to 0.056.

\subsection{RDCS\,1252.9--2927}

The FP in the $z=1.237$ cluster RDCS\,1252 was determined by
{Holden} {et~al.} (2005) using four galaxies, all of
which have $M>10^{11}\,M_{\odot}$.
Listed velocity dispersions are already on our system. The observed
$z_{850}$ surface brightnesses were converted from
$\langle \mu \rangle_e$ to $\mu_e$ and
converted to rest-frame $B$ using
\begin{equation}
B_z = z_{850} - 0.22 (i_{775} - J) + 0.99,
\end{equation}
where $z_{850}$, $i_{775}$, and $J$ are on the AB system
(see {Holden} {et~al.} 2005). This transformation is slightly different from
the one used in Holden et al.\ (2005), due to the use of
a different derivation procedure.\footnote{Note that most of
the apparent difference between Eq.\ A12 and the transformation
given in  Holden et al.\ (2005) results from the fact that in
their equation
the $z_{850}$ surface brightness is on the AB system and the
$i_{775}-J$ color is on the Vega system.}
For the typical colors of
galaxies in RDCS\,1252 the difference in the transformation
results in a difference of $\lesssim 0.05$\,mag in $B_z$.
The determination of velocity dispersions from the near-UV
spectral region may cause systematic errors of 5\,\%
({van Dokkum} \& {Stanford} 2003).
The total systematic uncertainty is 0.043 in $\log (M/L_B)$.
We note that for this cluster and for RDCS\,0848 the random uncertainty
exceeds the systematic uncertainty.

\subsection{RDCS\,0848+4453}

The FP for three galaxies in the $z=1.276$ cluster RDCS\,0848
was presented in {van Dokkum} \& {Stanford} (2003). The listed velocity dispersions
are on our system. The F160W surface brightnesses were corrected to
rest-frame $B$ using
\begin{equation}
B_z = H_{160} + 0.46 (I_{814}-H_{160}) + 1.92.
\end{equation}
This transformation is slightly different from
Eq.\ 3 in {van Dokkum} \& {Stanford} (2003), as we use a slightly
different synthetic $B$ filter in the present study.
For the typical colors of galaxies
in RDCS\,0848 the two transformations are consistent to $\lesssim 0.05$
mag. The systematic
uncertainty is a combination of the uncertainty derived above for
CL\,1358+62 and a 5\,\% uncertainty in the dispersions,
which originates from the use of the near-UV spectral region.
The adopted uncertainty in $\log (M/L_B)$ is 0.038.

\section{Literature Data for Field Galaxies}
\label{fielddata.sec}

\subsection{van Dokkum et al.\ (2001) and van Dokkum \& Ellis (2003)}

These studies give structural parameters and velocity dispersions for
morphologically-selected early-type galaxies. van Dokkum et al.\
(2001) give results for 18 galaxies at $0.15 \leq z \leq 0.55$
in the foreground of the rich clusters
MS\,2053--04 ($z=0.58$) and MS\,1054--03 ($z=0.83$).
van Dokkum \& Ellis studied ten galaxies at $0.56 \leq z \leq
1.10$ in the Hubble Deep Field North (HDF-N). The methodology
for deriving velocity dispersions and structural parameters
in these studies is identical to ours. Listed surface brightnesses
were transformed to rest-frame $B$ following the procedures
outlined in the cited papers. Note that this transformation is
different for each galaxy, as it depends on the redshift.
Effective radii were converted to kpc. We determined
masses and offsets
in $M/L_B$ ratio in the same way as for
individual cluster galaxies (see \S\,5.1). Eleven of 18
galaxies in the MS\,1054--03 field, and four of ten galaxies
in the HDF-N, have masses $>10^{11}$\,\msun.

\subsection{van der Wel et al.\ (2005)}

van der Wel et al.\ (2005) study a sample of 38 field
early-type galaxies in two
fields, the Chandra Deep Field South (CDF-S) and the field containing
the cluster RDCS\,1252$-29$ ($z=1.24$). The galaxies were selected
by a combination of color- and morphological criteria. They span
the redshift range
$0.62 \leq z \leq 1.14$, with median 0.97. van der Wel et al.\ follow
the same procedures as we do here for determining structural parameters
and velocity dispersions. The listed surface brightnesses
(Table 2 in van der Wel et al.\ 2005)
were transformed to the rest-frame $B$ band, and effective radii
were converted to kpc. Twenty-five galaxies in this sample
have $M>10^{11}$\,\msun.

\subsection{Treu et al.\ (2005a,b)}

Treu et al.\ study a very large sample of 226 visually classified
bulge-dominated galaxies in the GOODS HDF-N region. 
Although Treu et al.\ use different software and procedures than
we do here, a direct comparison between their data and those of van Dokkum
\& Ellis (2003) (which are on our system)
shows that any systematic differences are very small. Specifically,
for a subset of the galaxies velocity dispersions were determined
using the same software as used in the present study (see
\S\,2.5), and the two sets of measurements agree to $\lesssim 1$\,\%
on average (see Treu et al.\ 2005b).
Surface brightnesses were converted to rest-frame $B$ following
the procedure outlined in Treu et al.\ (2005b). Next, they
were converted from $\langle \mu \rangle_e$ to $\mu_e$, and
from the AB system to the Vega system. Effective
radii were converted to kpc. Treu et al.\ correct the measured
velocity dispersions for aperture effects by applying a fixed
correction of $\sigma_{\rm corr} = 1.10 \sigma_{\rm ap}$.
For consistency we correct the measured $\sigma_{\rm ap}$
values to the equivalent of a $3\farcs 4$ diameter
aperture at the distance of the Coma cluster (see \S\,2.5).

Masses and offsets in $M/L$ ratio were calculated following the
procedures of \S\,5.1. Of 141 galaxies classified as early-type
25 were excluded because they could be misclassified spirals,
as judged from
the residuals from an $r^{1/4}$ fit (see \S\,2.1
in Treu et al.\ 2005b). Of the remaining 116 galaxies, 48
have $M>10^{11}$\,\msun. These galaxies are at $0.32\leq z \leq 1.14$,
with median 0.84.

\subsection{Comparison of Individual Samples}

\begin{figure*}[t]
\epsfxsize=13cm
\epsffile[-180 195 511 637]{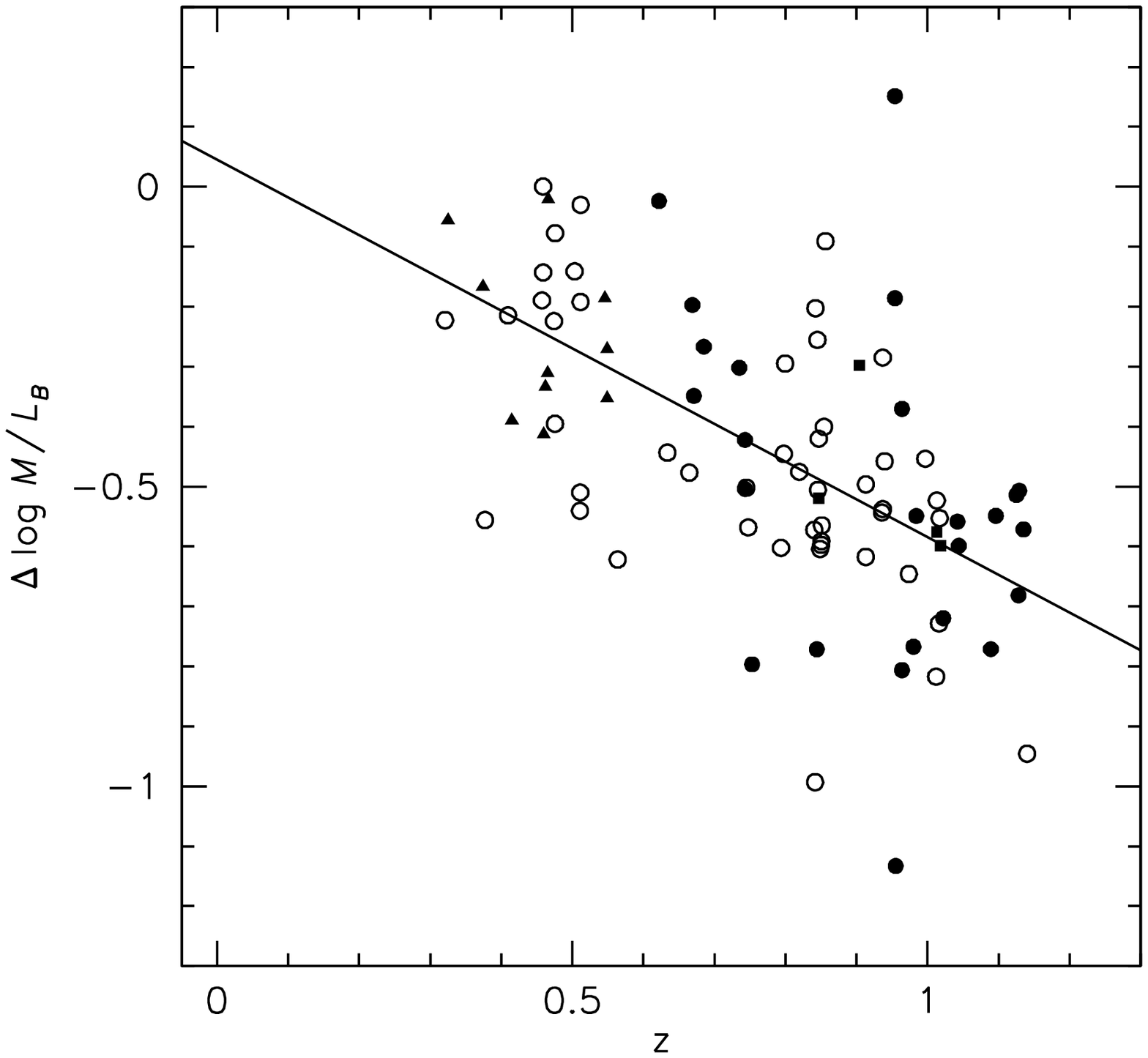}
\caption{\small
Individual field early-type galaxies with $M>10^{11}$\,\msun\
in the four samples that are combined in this study.
Filled triangles are galaxies from van Dokkum et al.\ (2001);
filled squares are from van Dokkum \& Ellis (2003); filled
circles are from van der Wel et al.\ (2005); and open
circles are from Treu et al.\ (2005b).
\label{mlind.plot}}
\end{figure*}

The total number of field early-type galaxies with $M>10^{11}$\,\msun\
is 88. One galaxy was excluded from this sample: object 1 from
van Dokkum et al.\ (2001). This galaxy is at $z=0.183$, and as
the next lowest redshift is $z=0.321$ including this object would
``artificially'' create a very wide bin at low redshift in Fig.\
12.\footnote{We note that this object follows the trend defined by the other 87
galaxies.} In \S\,6.2 we treat the remaining 87 galaxies
as a single sample; here we test whether this approach is
warranted, by examining whether there is evidence for
systematic differences between the four samples. Data points for
individual galaxies are shown in Fig.\ \ref{mlind.plot}.
We determined a mean offset for each sample in the following
way. We fitted a linear function to the binned field data
listed in Table 3. This function has the form $\Delta \log M/L_B
= 0.045 - 0.629 z$. After subtracting this linear function from
the individual data points, the average of the residuals was
determined for each of the four literature samples using the
biweight estimator. We find the following average offsets in
$\log M/L_B$: $-0.01 \pm 0.01$ (van Dokkum et al.\
2001); $-0.01 \pm 0.04$ (van Dokkum \& Ellis
2003); $+0.02 \pm 0.04$ (van der Wel et al.\
2005); and $-0.01 \pm 0.02$ (Treu et al.\ 2005b).
These values are consistent with each other, and
we conclude that it is appropriate to treat the four samples 
as a single, large sample.

\end{appendix}

\newpage

\begin{small}
\begin{center}
{ {\sc TABLE 2} \\
\sc Galaxy Sample} \\
\vspace{0.1cm}
\begin{tabular}{lccccrccl}
\hline
\hline
ID$^a$ & Type$^a$ & $z$ & $\sigma$\,(km/s)$^b$ & $\pm$ & $\log r_e\,(\arcsec)^c$ & $\mu_B$$^d$ &
$V_{555}-I_{814}$ & Remarks\\
\hline
\clusa--47 & E/S0 & $0.1308$ &  --- & --- & --- & --- & --- & Field galaxy$^e$ \\
\clusa--568 & S0/Sb & $0.4545$ & 152 & 10 & $-0.092$ & 23.68 & --- & Spiral galaxy$^{e}$\\
\clusa--834 & E & $0.4651$ & 165 & 14 & $-0.801$ & 21.35 & --- & \\
\clusa--868 & S0 & $0.4565$ & 197 & 18 & $-0.439$ & 22.93 & --- & \\
\clusa--968 & E & $0.4601$ & 213 & 12 & $-0.176$ & 23.64 & --- & \\
\clusa--2014 & E/S0 & $0.4616$ & 284 & 22 & $0.583$ & 25.33 & --- & AGN (3C\,295)\\
\clusbs--438 & E & $0.5399$ & 229 & 14 & $-0.269$ & 22.96 & 2.460 & \\
\clusbs--461 & E & $0.5458$ & 268 & 24 & $-0.532$ & 22.37 & 2.461 & \\
\clusbs--531 & E & $0.5420$ & 209 & 21 & $-0.687$ & 21.98 & 2.443 & \\
\clusbs--611 & E & $0.5509$ & 169 & 18 & $-0.441$ & 22.71 & 2.466 & \\
\clusbs--612 & E & $0.5508$ & 279 & 39 & $-0.322$ & 22.75 & 2.574 & \\
\clusbs--650 & E/S0 & $0.5445$ & 172 & 20 & $0.633$ & 26.35 & 2.482 & \\
\clusbs--659 & E & $0.5502$ & 264 & 27 & $-0.293$ & 23.13 & 2.441 & \\
\clusbs--724 & E & 0.5463 & 252 & 17 & 0.462 & 25.46 & 2.582 & \\
\clusbs--725 & E & 0.5389 & 226 & 18 & 0.739 & 26.18 & 2.531 & \\
\clusbs--745 & E & 0.5478 & 124 & 14 & $-0.105$ & 24.06 & 2.519 & \\
\clusbs--2050 & E & 0.5536 & 125 & 18 & $-0.444$ & 23.11 & 2.366 &  \\
\cluscs--270 & S0/E & 0.5098 & 186 & 12 & $-0.315$ & 22.84 & --- & Field early-type$^e$\\
\cluscs--292 & E & 0.5425 & 179 & 12 & $-0.301$ & 22.99 & --- & \\
\cluscs--474 & E/S0 & 0.5376 & 313 & 51 & 0.033 & 24.30 & ---& Merger$^e$\\
\cluscs--524 & E & 0.5403 & 226 & 16 & $-0.199$ & 23.50 & --- & \\
\cluscs--619 & E & 0.5392 & 232 & 12 & $-0.044$ & 24.00 & --- & \\
\cluscs--753 & E & 0.5395 &  281 & 12 & 0.131 & 24.20 & --- & \\
\cluscs--814 & E & 0.5409 & 226 & 19 & $-0.452$ & 22.88 & --- & \\
\cluscs--2040 & E & 0.5439 & 256 & 10 & $0.062$ & 23.83 & --- & \\
\cluscs--2043 & E & 0.5407 & 163 & 14 & $-0.461$ & 23.13 & --- & \\
\cluscs--2060 & E & 0.5430 & 235 & 36 & $-0.165$  & 23.43 & --- & \\
\hline
\end{tabular}
\end{center}
{\small
$^a$\,From Smail et al.\ (1997)\\
$^b$\,Velocity dispersion corrected to a $3\farcs 4$ diameter aperture at the
distance of the Coma cluster.\\
$^c$\,Circularized effective radius.\\
$^d$\,Surface brightness at the effective radius, in rest-frame $B$,
corrected for Galactic extinction but not corrected for cosmological
surface britghness dimming.\\
$^e$\,Not included in the analysis of the $M/L$ evolution.
}
\end{small}


\begin{references}

\reference{} Almeida, C., Baugh, C.~M., \& Lacey, C.~G.\ 2006, \mnras,
 submitted (astro-ph/0608544)

\reference{} {Baugh}, C.~M., {Cole}, S., \& {Frenk}, C.~S. 1996, \mnras, 283, 1361

\reference{} Baugh, C.~M., Lacey, C.~G., Frenk, C.~S., Granato, G.~L.,
  Silva, L., Bressan, A., Benson, A.~J., \& Cole, S.\ 2005, \mnras, 356, 1191

\reference{} {Beers}, T.~C., {Flynn}, K., \& {Gebhardt}, K. 1990, \aj, 100, 32

\reference{} {Bell}, E.~F., {Naab}, T., {McIntosh}, D.~H., {Somerville}, R.~S., {Caldwell},  J.~A.~R., {Barden}, M., {Wolf}, C., {Rix}, H.-W., {et al.} 2006, \apj, 640, 241

\reference{} {Bernardi}, M., {Nichol}, R.~C., {Sheth}, R.~K., {Miller}, C.~J., \&  {Brinkmann}, J. 2006, \aj, 131, 1288

\reference{} {Bernardi}, M., {Renzini}, A., {da Costa}, L.~N., {Wegner}, G., {Alonso},  M.~V., {Pellegrini}, P.~S., {Rit{\' e}}, C., \& {Willmer}, C.~N.~A. 1998,  \apjl, 508, L143

\reference{} {Bernardi}, M., {Sheth}, R.~K., {Annis}, J., {Burles}, S., {Eisenstein}, D.~J.,  {Finkbeiner}, D.~P., {Hogg}, D.~W., {Lupton}, R.~H., {et al.} 2003a, \aj, 125, 1866

\reference{} {Bernardi}, M., {Sheth}, R.~K., {Annis}, J., {Burles}, S., {Eisenstein}, D.~J.,  {Finkbeiner}, D.~P., {Hogg}, D.~W., {Lupton}, R.~H., {et al.} 2003b, \aj, 125, 1817

\reference{} Bessell, M.~S.\ 1990, \pasp, 102, 1181

\reference{} Blakeslee, J.~P., Holden, B.~P., Franx, M., Rosati, P.,
 Bouwens, R.~J., Demarco, R., Ford, H.~C., et al.\ 2006, \apj, 644, 30

\reference{} {Bower}, R.~G., {Benson}, A.~J., {Malbon}, R., {Helly}, J.~C., {Frenk}, C.~S.,  {Baugh}, C.~M., {Cole}, S., \& {Lacey}, C.~G. 2006, \mnras,
370, 645

\reference{} {Boylan-Kolchin}, M., {Ma}, C.-P., \& {Quataert}, E. 2006, \mnras, 534

\reference{} {Bruzual}, G. \& {Charlot}, S. 2003, \mnras, 344, 1000

\reference{} Buser, R., \& Kurucz, R.~L.\ 1978, A\&A, 70, 555

\reference{} {Cappellari}, M., {Bacon}, R., {Bureau}, M., {Damen}, M.~C., {Davies}, R.~L.,  {de Zeeuw}, P.~T., {Emsellem}, E., {Falc{\'o}n-Barroso}, J., {et al.} 2006, \mnras, 366, 1126

\reference{} {Chabrier}, G. 2003, \pasp, 115, 763

\reference{} {Clemens}, M.~S., {Bressan}, A., {Nikolic}, B., {Alexander}, P., {Annibali},  F., \& {Rampazzo}, R. 2006, \mnras, 649

\reference{} {Coleman}, G.~D., {Wu}, C.-C., \& {Weedman}, D.~W. 1980, \apjs, 43, 393

\reference{} Croton, D.~J, Springel, V., White, S.~D.~M., De Lucia, G.,
  Frenk, C.~S., Gao, L., Jenkins, A., Kauffmann, G., et al.\ 2006, \mnras,
  365, 11

\reference{} {Daddi}, E., {R{\"o}ttgering}, H.~J.~A., {Labb{\'e}}, I.,
 {Rudnick}, G., {Franx}, M., {Moorwood}, A.~F.~M., {Rix}, H.~W.,
 {van der Werf}, P.~P., \& {van Dokkum}, P.~G. 2003, \apj, 588, 50

\reference{} {Daddi}, E., {Cimatti}, A., {Renzini}, A., {Vernet}, J., {Conselice}, C.,  {Pozzetti}, L., {Mignoli}, M., {Tozzi}, P., {et al.} 2004, \apjl, 600, L127

\reference{} {De Lucia}, G., {Springel}, V., {White}, S.~D.~M., {Croton}, D., \&  {Kauffmann}, G. 2006, \mnras, 366, 499

\reference{} {Diaferio}, A., {Kauffmann}, G., {Balogh}, M.~L.,
{White}, S.~D.~M., {Schade}, D., \& {Ellingson}, E. 2001, \mnras, 323, 999

\reference{} {di Serego Alighieri}, S., {Lanzoni}, B., \& {J{\o}rgensen}, I. 2006, \apj, 647, L99

\reference{} {di Serego Alighieri}, S., {Vernet}, J., {Cimatti}, A., {Lanzoni}, B.,  {Cassata}, P., {Ciotti}, L., {Daddi}, E., {Mignoli}, M., {et al.} 2005, \aap,  442, 125

\reference{} {Djorgovski}, S. \& {Davis}, M. 1987, \apj, 313, 59

\reference{} {Dressler}, A., {Oemler}, A.~J., {Couch}, W.~J., {Smail}, I., {Ellis}, R.~S.,  {Barger}, A., {Butcher}, H., {Poggianti}, B.~M., {et al.} 1997,  \apj, 490, 577

\reference{} Faber, S.~M., Dressler, A., Davies, R.~L., Burstein, D.,
Lynden-Bell, D., Terlevich, R., Wegner, G.\ 1987, in Nearly Normal Galaxies,
ed.\ S.~M.\ Faber (New York: Springer), 175

\reference{} Ferreras, I., Lisker, T., Carollo, C.~M., Lilly, S.~J.,
 Mobasher, B.\ 2005, \apj, 635, 243

\reference{} {F{\"o}rster Schreiber}, N.~M., {van Dokkum}, P.~G., {Franx}, M., {Labb{\'e}},  I., {Rudnick}, G., {Daddi}, E., {Illingworth}, G.~D., {Kriek}, M., {et al.} 2004, \apj, 616, 40

\reference{} {Franx}, M. 1993, \pasp, 105, 1058

\reference{} {Franx}, M., {Labb{\' e}}, I., {Rudnick}, G., {van Dokkum}, P.~G., {Daddi}, E.,  {F{\" o}rster Schreiber}, N.~M., {Moorwood}, A., {Rix}, H., {et al.} 2003, \apjl, 587, L79

\reference{} Freedman, W.~L., Madore, B.~F., Gibson, B.~K., Ferrarese, L.,
Kelson, D.~D., Sakai, S., Mould, J.~R., et al.\ 2001, \apj, 553, 47

\reference{} {Fritz}, A., {Ziegler}, B.~L., {Bower}, R.~G., {Smail}, I., \& {Davies}, R.~L.  2005, \mnras, 358, 233

\reference{} {Fukugita}, M., {Ichikawa}, T., {Gunn}, J.~E., {Doi}, M., {Shimasaku}, K., \&  {Schneider}, D.~P. 1996, \aj, 111, 1748

\reference{} {G{\'o}mez}, P.~L., {Hughes}, J.~P., \& {Birkinshaw}, M. 2000, \apj, 540, 726

\reference{} {Gonz{\' a}lez-Garc{\'{\i}}a}, A.~C. \& {van Albada}, T.~S. 2003, \mnras, 342,  L36

\reference{} {Grazian}, A., {Fontana}, A., {Moscardini}, L., {Salimbeni}, S.,
 {Menci}, N., {Giallongo}, E., {de Santis}, C., {Gallozzi}, S., et al.
 2006, A\&A, 453, 507

\reference{} {Guzman}, R., {Koo}, D.~C., {Faber}, S.~M., {Illingworth}, G.~D., {Takamiya},  M., {Kron}, R.~G., \& {Bershady}, M.~A. 1996, \apjl, 460, L5

\reference{} {Holden}, B.~P., {van der Wel}, A., {Franx}, M., {Illingworth}, G.~D.,  {Blakeslee}, J.~P., {van Dokkum}, P., {Ford}, H., {Magee}, D., {et al.} 2005, \apjl, 620, L83
	
\reference{} Holden, B.~P., Franx, M., Illingworth, G.~D., Postman, M.,
Blakeslee, J.~P., Homeier, N., Demarco, R., et al.\ 2006, \apj, 642, L123

\reference{} {Holtzman}, J.~A., {Burrows}, C.~J., {Casertano}, S., {Hester}, J.~J.,  {Trauger}, J.~T., {Watson}, A.~M., \& {Worthey}, G. 1995, \pasp, 107, 1065

\reference{} {Hudson}, M.~J., {Lucey}, J.~R., {Smith}, R.~J., {Schlegel}, D.~J., \&  {Davies}, R.~L. 2001, \mnras, 327, 265

\reference{} J\o{}rgensen, I., Bergmann, M., Davies, R., Barr, J.,
  Takamiya, M., \& Crampton, D.\ 2005, \aj, 129, 1249

\reference{} {J{\o}rgensen}, I., {Chiboucas}, K., {Flint}, K., {Bergmann}, M., {Barr}, J., \&  {Davies}, R. 2006, \apjl, 639, L9
	
\reference{} {J{\o}rgensen}, I., {Franx}, M., {Hjorth}, J., \& {van Dokkum}, P.~G. 1999,  \mnras, 308, 833

\reference{} {J{\o}rgensen}, I., {Franx}, M., \& {Kj\ae{}rgaard}, P. 1992, \aaps, 95, 489

\reference{} J{\o}rgensen, I., {Franx}, M., \& {Kj\ae{}rgaard}, P. 1995a, \mnras, 273, 1097

\reference{} {J{\o}rgensen}, I., {Franx}, M., \& {Kj\ae{}rgaard}, P. 1995b, \mnras, 276, 1341

\reference{} ---. 1996, \mnras, 280, 167

\reference{} {Kang}, X., {Jing}, Y.~P., \& {Silk}, J. 2006, ApJ, 648,
820

\reference{} {Kauffmann}, G. 1996, \mnras, 281, 487

\reference{} {Kauffmann}, G. \& {Charlot}, S. 1998, \mnras, 297, L23

\reference{} {Kauffmann}, G., {Colberg}, J.~M., {Diaferio}, A., \&
{White}, S.~D.~M. 1999, \mnras, 303, 188

\reference{} {Kelson}, D.~D. 2003, \pasp, 115, 688

\reference{} {Kelson}, D.~D., {Illingworth}, G.~D., {van Dokkum}, P.~G., \& {Franx}, M.  2000a, \apj, 531, 137

\reference{} ---. 2000b, \apj, 531, 159

\reference{} ---. 2000c, \apj, 531, 184

\reference{} {Kelson}, D.~D., {van Dokkum}, P.~G., {Franx}, M., {Illingworth}, G.~D., \&  {Fabricant}, D. 1997, \apjl, 478, L13

\reference{} {Knudsen}, K.~K., {van der Werf}, P., {Franx}, M., {F{\"o}rster Schreiber},  N.~M., {van Dokkum}, P.~G., {Illingworth}, G.~D., {Labb{\'e}}, I.,  {Moorwood}, A., {et al.} 2005, \apjl, 632, L9

\reference{} {Kriek}, M., {van Dokkum}, P., {Franx}, M., {Forster Schreiber}, N., {Gawiser},  E., {Illingworth}, G., {Labbe}, I., {Marchesini}, D., {et al.} 2006,  ApJ, 645, 44

\reference{} Kriek, M., van Dokkum, P., Franx, M., Quadri, R., Gawiser, E.,
 Herrera, D., Illingworth, G., Labb\'e, I., et al.\ 2006, \apjl, in press
  (astro-ph/0608446)

\reference{} {Krist}, J. 1995, in ASP Conf. Ser. 77: Astronomical Data Analysis Software and  Systems IV, Vol.~4, 349

\reference{} {Kroupa}, P. 2001, \mnras, 322, 231

\reference{} {Labb{\' e}}, I., {Franx}, M., {Rudnick}, G., {Schreiber}, N.~M.~F., {Rix}, H.,  {Moorwood}, A., {van Dokkum}, P.~G., {van der Werf}, P., {et al.} 2003, \aj, 125, 1107

\reference{} {Labb{\'e}}, I., {Huang}, J., {Franx}, M., {Rudnick}, G., {Barmby}, P.,  {Daddi}, E., {van Dokkum}, P.~G., {Fazio}, G.~G., {et al.} 2005, \apjl, 624, L81

\reference{} {Larson}, R.~B. 1998, \mnras, 301, 569

\reference{} {Le Borgne}, J.~F., {Pello}, R., \& {Sanahuja}, B. 1992, \aaps, 95, 87

\reference{} {Lubin}, L.~M. \& {Sandage}, A. 2001, \aj, 122, 1084

\reference{} {Maraston}, C. 2005, \mnras, 362, 799

	
\reference{} Mei, S., Holden, B.~P., Blakeslee, J.~P., Rosati, P.,
Postman, M., Jee, M.~J., Rettura, A., et al.\ 2006, \apj, 644, 759

\reference{} {Moran}, S.~M., {Ellis}, R.~S., {Treu}, T., {Smail}, I., {Dressler}, A.,  {Coil}, A.~L., \& {Smith}, G.~P. 2005, \apj, 634, 977

\reference{} {Nagamine}, K., {Cen}, R., {Hernquist}, L., {Ostriker}, J.~P., \& {Springel},  V. 2005, \apj, 627, 608

\reference{} {Nagashima}, M., {Lacey}, C.~G., {Okamoto}, T., {Baugh}, C.~M., {Frenk}, C.~S.,  \& {Cole}, S. 2005, \mnras, 363, L31

\reference{} {Pahre}, M.~A., {Djorgovski}, S.~G., \& {de Carvalho}, R.~R. 1996, \apjl, 456,  L79+

\reference{} {Papovich}, C., {Moustakas}, L.~A., {Dickinson}, M., {Le Floc'h}, E., {Rieke},  G.~H., {Daddi}, E., {Alexander}, D.~M., {Bauer}, F., {et al.} 2006, \apj, 640, 92

\reference{} {Quadri}, R., {van Dokkum}, P., {Gawiser}, E., {Franx}, M.,
 {Marchesini}, D., {Lira}, P., {Rudnick}, G., {Herrera}, D., et al. 2006,
 ApJ, in press (astro-ph/0606330)

\reference{} {Reddy}, N.~A., {Erb}, D.~K., {Steidel}, C.~C., {Shapley}, A.~E., {Adelberger},  K.~L., \& {Pettini}, M. 2005, \apj, 633, 748

\reference{} {Robertson}, B., {Cox}, T.~J., {Hernquist}, L., {Franx}, M., {Hopkins}, P.~F.,  {Martini}, P., \& {Springel}, V. 2006, \apj, 641, 21

\reference{} {Rubin}, K.~H.~R., {van Dokkum}, P.~G., {Coppi}, P., {Johnson}, O.,  {F{\"o}rster Schreiber}, N.~M., {Franx}, M., \& {van der Werf}, P. 2004,  \apjl, 613, L5

\reference{} Rusin, D., Kochanek, C.~S., Falco, E.~E., Keeton, C.~R.,
  McLeod, B.~A., Impey, C.~D., Leh\'ar, J., et al.\ 2003, \apj, 587, 143

\reference{} Rusin, D., \& Kochanek, C.~S.\ 2005, \apj, 623, 666

\reference{} {Salpeter}, E.~E. 1955, \apj, 121, 161

\reference{} {Schlegel}, D.~J., {Finkbeiner}, D.~P., \& {Davis}, M. 1998, \apj, 500, 525

\reference{} {Schneider}, R., {Ferrara}, A., {Natarajan}, P., \& {Omukai}, K. 2002, \apj,  571, 30

\reference{} Sirianni, M.~J.~J., Ben\'\i{}tez, N., Blakeslee, J.~P.,
 Martel, A.~R., Meurer, G., Clampin, M., de Marchi, G., Ford, H.~C.,
 et al.\ 2005, \pasp, 117, 1049

\reference{} {Smail}, I., {Dressler}, A., {Couch}, W.~J., {Ellis}, R.~S., {Oemler}, A.~J.,  {Butcher}, H., \& {Sharples}, R.~M. 1997, \apjs, 110, 213 [MORPHS]

\reference{} {Steidel}, C.~C., {Adelberger}, K.~L., {Shapley}, A.~E., {Erb}, D.~K., {Reddy},  N.~A., \& {Pettini}, M. 2005, \apj, 626, 44

\reference{} {Thomas}, D., {Maraston}, C., {Bender}, R., \& {Mendes de Oliveira}, C. 2005,  \apj, 621, 673

\reference{} {Tinsley}, B.~M. 1980, Fundamentals of Cosmic Physics, 5, 287

\reference{} {Tonry}, J. \& {Davis}, M. 1979, \aj, 84, 1511

\reference{} {Tran}, K.-V.~H., {van Dokkum}, P., {Franx}, M., {Illingworth}, G.~D.,  {Kelson}, D.~D., \& {Schreiber}, N.~M.~F. 2005a, \apjl, 627, L25

\reference{} {Tran}, K.-V.~H., {van Dokkum}, P., {Illingworth}, G.~D., {Kelson}, D.,  {Gonzalez}, A., \& {Franx}, M. 2005b, \apj, 619, 134

\reference{} {Treu}, T., {Ellis}, R.~S., {Liao}, T.~X., \& {van Dokkum}, P.~G.  2005a, \apjl, 622, L5

\reference{} {Treu}, T., {Ellis}, R.~S., {Liao}, T.~X., {van Dokkum}, P.~G., {Tozzi}, P.,  {Coil}, A., {Newman}, J., {Cooper}, M.~C., {et al.} 2005b,  \apj, 633, 174

\reference{} {Treu}, T., {Stiavelli}, M., {Casertano}, S., {Moller}, P., \& {Bertin}, G.  1999, \mnras, 308, 1037

\reference{} ---. 2002, \apjl, 564, L13

\reference{} van der Marel, R.~P., \& van Dokkum, P.~G. 2006a, \apj, submitted [vdMvD06a] (astro-ph/0611571)

\reference{} van der Marel, R.~P., \& van Dokkum, P.~G. 2006b, \apj, submitted [vdMvD06b] (astro-ph/0611577)

\reference{} {van der Wel}, A., {Franx}, M., {van Dokkum}, P.~G., \& {Rix}, H.-W. 2004,  \apjl, 601, L5

\reference{} {van der Wel}, A., {Franx}, M., {van Dokkum}, P.~G., {Rix}, H.-W.,  {Illingworth}, G.~D., \& {Rosati}, P. 2005, \apj, 631, 145

\reference{} van de Ven, G., van Dokkum, P.~G., \& Franx, M.\ 2003, \mnras,
 344, 924

\reference{} {van Dokkum}, P.~G. 2001, \pasp, 113, 1420

\reference{} ---. 2005, \aj, 130, 2647

\reference{} {van Dokkum}, P.~G. \& {Ellis}, R.~S. 2003, \apjl, 592, L53

\reference{} {van Dokkum}, P.~G., {F{\" o}rster Schreiber}, N.~M., {Franx}, M., {Daddi}, E.,  {Illingworth}, G.~D., {Labb{\' e}}, I., {Moorwood}, A., {Rix}, H., {et al.} 2003, \apjl, 587, L83

\reference{} {van Dokkum}, P.~G. \& {Franx}, M. 1996, \mnras, 281, 985

\reference{} ---. 2001, \apj, 553, 90

\reference{} {van Dokkum}, P.~G., {Franx}, M., {Kelson}, D.~D., \& {Illingworth}, G.~D.  1998a, \apjl, 504, L17

\reference{} ---. 2001, \apjl, 553, L39

\reference{} {van Dokkum}, P.~G., {Franx}, M., {Kelson}, D.~D., {Illingworth}, G.~D.,  {Fisher}, D., \& {Fabricant}, D. 1998b, \apj, 500, 714

\reference{} {van Dokkum}, P.~G., {Quadri}, R., {Marchesini}, D., {Rudnick}, G., {Franx},  M., {Gawiser}, E., {Herrera}, D., {Wuyts}, S., {et al.} 2006, \apjl, 638, L59

\reference{} {van Dokkum}, P.~G. \& {Stanford}, S.~A. 2003, \apj, 585, 78

\reference{} {Webb}, T.~M.~A., {van Dokkum}, P., {Egami}, E., {Fazio}, G., {Franx}, M.,  {Gawiser}, E., {Herrera}, D., {Huang}, J., {et al.} 2006, \apjl, 636, L17

\reference{} {Worthey}, G. 1994, \apjs, 95, 107

\reference{} {Worthey}, G., {Faber}, S.~M., \& {Gonzalez}, J.~J. 1992, \apj, 398, 69

\reference{} {Wuyts}, S., {van Dokkum}, P.~G., {Kelson}, D.~D., {Franx}, M., \&  {Illingworth}, G.~D. 2004, \apj, 605, 677

\reference{} {Yan}, H., {Dickinson}, M., {Eisenhardt}, P.~R.~M., {Ferguson}, H.~C.,  {Grogin}, N.~A., {Paolillo}, M., {Chary}, R.-R., {Casertano}, S., {et al.} 2004, \apj, 616, 63

\reference{} {Ziegler}, B.~L., {Bower}, R.~G., {Smail}, I., {Davies}, R.~L., \& {Lee}, D.  2001, \mnras, 325, 1571

\end{references}
\end{document}